\documentclass[letterpaper,draftclsnofoot,onecolumn,oneside,11pt]{IEEEtran}
\usepackage[noadjust]{cite}
\usepackage{latexsym}
\usepackage[latin1]{inputenc}
\usepackage[T1]{fontenc}
\usepackage{graphicx}
\usepackage{verbatim,perso}
\usepackage{latexsym}
\usepackage{stfloats}
\usepackage{float}
\usepackage{multirow}
\usepackage{subfigure}
\usepackage{stackengine}
\usepackage{amsmath,amsbsy,amsfonts,amssymb}
\usepackage{mathtools}
\mathtoolsset{showonlyrefs}
\usepackage{bbm,bm}
\usepackage{relsize}
\usepackage{kbordermatrix}
\renewcommand{\kbldelim}{(}
\renewcommand{\kbrdelim}{)}

\usepackage{bigstrut}
\usepackage{psfrag}
\usepackage{blkarray}
\usepackage{tikz}
\usetikzlibrary{automata,positioning,calc}
\usetikzlibrary{shapes,arrows}
\usepackage{xcolor}
\allowdisplaybreaks

\usepackage{lipsum}

\usepackage[nolist]{acronym}

\newcommand{\mc}{\mathcal}
\newcommand{\ms}{\mathsf}

\DeclareMathOperator{\rank}{rank}

\newfloat{EqAtPageBottom}{b}{}

\usepackage[colorinlistoftodos,prependcaption,textsize=footnotesize]{todonotes}

\title{Multiple-Relay Slotted ALOHA: \\ Performance Analysis and Bounds}
\author{
\vspace{3mm}
\normalsize{Andrea Munari,~\IEEEmembership{\normalsize Senior Member,~IEEE},
Federico Clazzer,~\IEEEmembership{\normalsize Member,~IEEE}, \\Gianluigi Liva,~\IEEEmembership{\normalsize Senior Member,~IEEE}, Michael Heindlmaier}
\thanks{A. Munari, F. Clazzer and G. Liva are with the Inst.\ of Communications and Navigation, German Aerospace Center (DLR), Oberpfaffenhofen, Germany (e-mail: \{andrea.munari,federico.clazzer,gianluigi.liva\}@dlr.de),\newline
M. Heindlmaier was with the Inst. for Communications Engineering, Technische Universit\"{a}t M\"{u}nchen, Theresienstr. 90, 80333 Munich, Germany (e-mail: michael.heindlmaier@tum.de). He is now with Cadami GmbH, M\"{u}nchen. \newline
Part of this work has been presented at the 51$^{\mbox{\tiny st}}$ Annual Allerton Conference on Communication, Control, and Computing, Oct. 2013.
}
}

\date{}

\begin{document}

\begin{acronym}
\acro{AWGN}{additive white gaussian noise}
\acro{ACRDA}{asynchronous contention resolution diversity ALOHA}
\acro{CDF}{cumulative distribution function}
\acro{CoMP}{coordinated multi-point}
\acro{CRA-CC}{CRA-convolutional code}
\acro{CRA-SH}{CRA-shannon bound}
\acro{CRA}{contention resolution ALOHA}
\acro{CRDSA}{contention resolution diversity slotted ALOHA}
\acro{CRDSA++}{contention resolution diversity slotted ALOHA++}
\acro{CSA}{coded slotted ALOHA}
\acro{DAMA}{demand assigned multiple access}
\acro{DSA}{diversity slotted ALOHA}
\acro{ECRA}{enhanced contention resolution ALOHA}
\acro{ECRA-SC}{ECRA selection combining}
\acro{ECRA-MRC}{ECRA maximal-ratio combining}
\acro{EGD}{equal-gain diversity}
\acro{ETP}{equal transmission power}
\acro{FEC}{forward error correction}
\acro{GEO}{geostationary orbit}
\acro{GW}{gateway}
\acro{IC}{interference cancellation}
\acro{IoT}{internet of things}
\acro{IRCRA}{irregular repetition contention resolution ALOHA}
\acro{IRSA}{irregular repetition slotted ALOHA}
\acro{M2M}{machine-to-machine}
\acro{MAC}{medium access}
\acro{MF-TDMA}{multi-frequency time division multiple access}
\acro{MIMO}{multiple-input multiple-output}
\acro{MPR}{multi-packet reception}
\acro{MRC}{maximal-ratio combining}
\acro{MTC}{machine-type communications}
\acro{OOF}{on-off fading}
\acro{PDF}{probability density function}
\acro{PER}{packet error rate}
\acro{PLR}{packet loss rate}
\acro{PMF}{probability mass function}
\acro{PPC}{perfect power control}
\acro{RA}{Random access}
\acro{RCB}{random coding bound}
\acro{RLC}{random linear coding}
\acro{RTT}{round trip time}
\acro{SA}{slotted ALOHA}
\acro{SB}{Shannon bound}
\acro{SC}{selection combining}
\acro{SFP}{simplified forwarding policy}
\acro{SIC}{successive interference cancellation}
\acro{SIR}{signal to interference ratio}
\acro{SNIR}{signal-to-noise and interference ratio}
\acro{SINR}{signal-to-interference and noise ratio}
\acro{SNR}{signal-to-noise ratio}
\acro{TDMA}{time division multiple access}
\acro{UCP}{unresolvable collision pattern}
\end{acronym}
\newtheorem{thm}{Theorem}
\newtheorem{lemma}{Lemma}
\newtheorem{coroll}{Corollary}
\newtheorem{prop}{Proposition}
\newtheorem{result}{Result}
\newtheorem{conj}{Conjecture}
\newtheorem{example}{Example}

\newcommand{\pr}{\ensuremath{\mathbb P}}

\newcommand{\oleq}[1]{\overset{\text{(#1)}}{\leq}}
\newcommand{\oeq}[1]{\overset{\text{(#1)}}{=}}
\newcommand{\ogeq}[1]{\overset{\text{(#1)}}{\geq}}
\newcommand{\ogeql}[2]{\overset{#1}{\underset{#2}{\gtreqless}}}

\newcommand{\load}{\ensuremath{\mathsf{G}}}

\newcommand{\tp}{\mathsf{S}}
\newcommand{\tpSA}{\tp_{\mathsf{sa}}}
\newcommand{\tpUL}{\ensuremath{\tp_{\ulVal}}}
\newcommand{\tpMax}{\tp^{*}}
\newcommand{\tpULMax}{\tpMax_{\ulVal}}
\newcommand{\tpDL}{\tp_{\dlVal}}

\newcommand{\plrUL}{\ensuremath{\mathsf p_{\ulVal}}}

\newcommand{\peras}{\ensuremath{\varepsilon}}

\newcommand{\slotTime}{\ensuremath{\mathsf{T}}}

\newcommand{\slot}{m}
\newcommand{\usersRV}{\ensuremath{U}}
\newcommand{\users}{\ensuremath{u}}
\newcommand{\usInd}{\ensuremath{\nu}}
\newcommand{\rxSet}{\ensuremath{\mc K}}
\newcommand{\nrx}{\ensuremath{\mathsf{K}}}
\newcommand{\rx}{k}
\newcommand{\setArx}{\ensuremath{\mathcal{A}_\rx}}
\newcommand{\rxSubSet}{\ensuremath{\mathcal{J}}}
\newcommand{\nrxSubSet}{\ensuremath{\mathsf{J}}}
\newcommand{\subsetrx}{j}
\newcommand{\setAsubset}{\ensuremath{\mathcal{A}_\subsetrx}}
\newcommand{\ulVal}{\ensuremath{\mathsf{ul}}}
\newcommand{\ulslot}{\ensuremath{\mathsf{T_{\ulVal}}}}
\newcommand{\collRV}{\ensuremath{C}}
\newcommand{\collrv}{\ensuremath{c}}
\newcommand{\interfRV}{\ensuremath{I}}
\newcommand{\interfrv}{\ensuremath{i}}
\newcommand{\numulslot}{\ensuremath{m_{\ulVal}}}
\newcommand{\psamecoll}{\ensuremath{\phi_\nrxSubSet}}
\newcommand{\dlVal}{\ensuremath{\mathsf{dl}}}
\newcommand{\dlslot}{\ensuremath{\mathsf{T_{\dlVal}}}}
\newcommand{\numdlslot}{\ensuremath{m_{\dlVal}}}
\newcommand{\dlrate}{\ensuremath{\mathsf{R}}}
\newcommand{\dlratemin}{\ensuremath{\dlrate^*}}
\newcommand{\dlratek}{\ensuremath{\mathsf{R}_k}}
\newcommand{\dltxk}{\ensuremath{z_k}}
\newcommand{\plos}{\ensuremath{\mathsf p_{\dlVal}}}

\newcommand{\dummy}{\ensuremath{\ell}}
\newcommand{\payload}{\ensuremath{L}}
\newcommand{\erasure}{\ensuremath{e}}
\newcommand{\fieldOrd}{q}

\newcommand{\linll}{w}
\newcommand{\linllK}[1]{\linll_{#1}}
\newcommand{\linComb}{\bm{\linll}}
\newcommand{\linCombK}[1]{\linComb_{#1}}
\newcommand{\linCombRed}{\tilde{\bm{w}}}
\newcommand{\linCombRedK}[1]{\linCombRed_{#1}}
\newcommand{\MxComb}{\bm{G}}
\newcommand{\MxCombK}[1]{\MxComb_{#1}}
\newcommand{\MxCombRed}{\tilde{\MxComb}}
\newcommand{\MxCombRedK}[1]{\MxCombRed_{#1}}
\newcommand{\coll}{v}
\newcommand{\collMax}{\coll^{*}}
\newcommand{\collK}[1]{\coll_{#1}}
\newcommand{\colVec}{\bm{\coll}}
\newcommand{\colVecK}[1]{\colVec_{#1}}
\newcommand{\colVecNum}{\bar{\colVec}}

\newcommand{\USet}{\ensuremath{\mathcal{U}}}
\newcommand{\LSet}{\ensuremath{\mathcal{L}}}
\newcommand{\qij}[2]{\ensuremath{q_{ #1,#2 } } }
\newcommand{\dropij}[2]{\ensuremath{\chi_{ #1,#2 } } }
\newcommand{\dropVec}{\ensuremath{\bm \chi } }
\newcommand{\drop}{\ensuremath{\chi } }
\newcommand{\arrRate}[1]{\ensuremath{a_{#1} } }
\newcommand{\coeffA}{\ensuremath{\Delta}}
\newcommand{\coeffB}{\ensuremath{\Theta}}
\newcommand{\coeffC}{\ensuremath{\Upsilon}}
\newcommand{\coeffD}{\ensuremath{\Psi}}
\newcommand{\coeffE}{\ensuremath{\Lambda}}

\newcommand{\capDL}{\mathsf{\overline{S}_{\mathsf{dl}}}}
\newcommand{\Id}{\ensuremath{\bm{I}}}
\newcommand{\MxA}{\ensuremath{\bm{A}}}
\newcommand{\MxB}{\ensuremath{\bm{B}}}
\newcommand{\MxUone}{\ensuremath{\bm{U}_1}}
\newcommand{\MxUtwo}{\ensuremath{\bm{U}_2}}
\newcommand{\MxL}{\ensuremath{\bm{L}}}

\newcommand{\rankk}{n}
\newcommand{\rankK}[1]{\rankk_{#1}}

\newcommand{\mxCol}{h}
\newcommand{\mxColK}[1]{\mxCol_{#1}}

\newcommand{\collSinkone}{\ensuremath{d_1}}
\newcommand{\collSinktwo}{\ensuremath{d_2}}
\newcommand{\collSinkboth}{\ensuremath{d_{1,2}}}
\newcommand{\collSinkoneRV}{\ensuremath{D_1}}
\newcommand{\collSinktwoRV}{\ensuremath{D_2}}
\newcommand{\collSinkbothRV}{\ensuremath{D_{1,2}}}

\newcommand{\MxM}{\ensuremath{\bm{M}}}

\newcommand{\state}{\ensuremath{\bm{x}}}

\newcommand{\eo}{\ensuremath{\bm{e}_1}}
\newcommand{\et}{\ensuremath{\bm{e}_2}}
\newcommand{\etr}{\ensuremath{\bm{e}_3}}
\newcommand{\Tx}{\ensuremath{\psi}}

\newcommand{\st}{\ensuremath{\pi}}
\newcommand{\x}{\mathtt{x}}
\newcommand{\hSet}{\ensuremath{\mc S}}
\newcommand{\vSet}{\ensuremath{\mc V}}
\newcommand{\sv}{\ensuremath{\xi}}

\newcommand{\plrI}{\ensuremath{\mathsf P_{\mathsf{L}|\mathsf I}}}

\definecolor{gl}{rgb}{0.0,0.5,0.8}
\definecolor{fc}{rgb}{0.8,0.5,0}
\definecolor{al}{rgb}{1,0.3,0.3}
\newcommand{\giangio}{\textcolor{gl}}
\newcommand{\fede}{\textcolor{fc}}
\newcommand{\alert}{\textcolor{al}}

\setcounter{page}{1}
\maketitle
\thispagestyle{empty}

\begin{abstract}
Wireless random access protocols are attracting a revived research interest as a simple yet effective solution for machine-type communications. In the quest to improve reliability and spectral efficiency of such schemes, the use of multiple receivers has recently emerged as a promising option. We study the potential of this approach considering a population of users that transmit data packets following a simple slotted ALOHA policy to a set of non-cooperative receivers or relays (\emph{uplink} phase). These, in turn, independently forward \--- part of \--- what decoded towards a collecting sink (\emph{downlink} phase). For an on-off fading channel model, we provide exact expressions for uplink throughput and packet loss rate for an arbitrary number of relays, characterising the benefits of multi-receiver schemes. Moreover, a lower bound on the minimum amount of downlink resources needed to deliver all information collected on the uplink is provided. The bound is proven to be achievable via random linear coding when no constraints  in terms of latency are set. We complement our study discussing a family of simple forwarding policies that require no packet-level coding, and optimising their performance based on the amount of available downlink resources.
The behaviour of both random linear coding and simplified policies is also characterised when receivers are equipped with finite buffers, revealing non-trivial tradeoffs.
\end{abstract}


\section{Introduction} \label{sec:introduction}
%

\IEEEPARstart{I}{nterest} in multiple access protocols for wireless networks has steadily gained momentum in the past few years, thanks to the rise of the \ac{MTC} paradigm. New applications blooming in this domain are characterised by the presence of a massive population of terminals \--- often with limited capabilities in terms of hardware or power \--- sharing a common channel for sporadic transmissions of short data packets. In view of such features, \ac{MTC} pose new and unique challenges that span multiple layers, calling for design principles that often depart from those of traditional and well-established wireless systems. From the medium access standpoint, in particular, schedule-based solutions are largely inefficient when small amounts of data are to be transferred in an unpredictable fashion, due to the large overhead needed to coordinate resource allocation. \ac{RA} policies appear instead especially appealing, and slightly modified versions of ALOHA \cite{Abramson:ALOHA} have already made their way to commercial solutions, e.g. \cite{LoRa,SigFox,Dai15}.
The performance of such schemes is however inherently limited by collisions \cite{Abramson77:PacketBroadcasting}, making them unsuitable to fully support the high throughput or stringent reliability requirements encountered in many relevant \ac{MTC} settings \cite{Popovski18_Network}.

To bridge this gap, a revived attention for \ac{RA} has led to the development of protocols that apply the principle of \emph{diversity} to ALOHA. A first and flourishing line of research in this direction combines the original idea of Choudhury and Rappaport \cite{Rappaport83:DSA}, i.e. having each transmitter proactively send multiple copies of a packet over time, with the use of \ac{SIC} at the receiver \cite{DeGaudenzi07:CRDSA, Liva11:IRSA, Narayan12:BoundIRSA, Stefanovic13:RatelessAloha, DeGaudenzi14:ACRDA, Paolini15:TIT_CSA, Sandgren17:FACSA, Polyanskiy17:RA, Ordentlich17:RA, Clazzer18:ECRA}. Specifically, whenever a data unit is decoded, interference generated by its twins can be removed, possibly rendering other previously collided packets retrievable. A deep understanding of the behaviour of such protocols has been achieved borrowing tools of codes on graphs \cite{Richardson01_LDPC}, showing how an accurate design of the probability mass function used by nodes to draw the number of sent replicas allows to approach the ultimate $1$ $\mathrm{[pk/slot]}$ throughput limit of a collision channel for an unbounded delay \cite{Narayan12:BoundIRSA}. Building on this idea, several protocols have been devised, spanning from the time synchronous \ac{CSA} \cite{Liva11:IRSA,Paolini15:TIT_CSA} and frameless ALOHA \cite{Stefanovic13:RatelessAloha}, to asynchronous alternatives \cite{Clazzer18:ECRA,DeGaudenzi14:ACRDA}, some of which have been embraced by current standards~\cite{dvbrcs2}.

Despite these remarkable results, solutions that rely on \emph{time diversity} require modifications at the transmitter side compared to the use of plain ALOHA, entailing additional complexity and possibly hindering their seamless application to already deployed systems. In view of this, research efforts have also focused on the potential of \emph{spatial diversity} for \ac{RA}, studying settings in which nodes transmit over the wireless medium a single copy of their data units, whose reception is attempted at different positions. Along this line, Zorzi and LaMaire \cite{Zorzi98:ALOHA,LaMaire96:Diversity} characterised the performance of ALOHA when a single receiver is equipped with multiple properly spaced antennas, and propagation is affected by shadowing and fading. Results were further extended from an information theoretic viewpoint by Tse et al. \cite{Tse04:DivMAC}, considering the possibility to jointly process incoming signals at the receiver's antennas.

Spatial diversity can also be leveraged without resorting to multi-antenna terminals, and instead having a set of disjoint receivers attempt collection of data packets transmitted over the shared medium. The idea was pioneered by Corson and Ephremides in the early $1990$s \cite{Ephremides93_MultiReceiver}, and was recently revived in cellular scenarios considering uplink reception at multiple base stations \cite{Jakovetic15:TCOM, Ogata15:FramelessMulRec, Ogata17:CoopFrameless, Mastilovic17:CoopSADirAntenna}.\footnote{Incidentally, we note that the potential of spatial diversity has been studied for scheduled uplink access as well, see, e.g. \cite{Simeone09:CoopUp}, and is leveraged in LTE-A with \ac{CoMP} \cite{Bassoy17_CoMP}.} Specifically, these works consider once more variations of ALOHA based on time repetition \--- thus combining \emph{spatial} and \emph{time} diversity \--- and aim at optimising the probability mass function employed by transmitters to send their replicas assuming different degrees of coordination among the receivers. Remarkable throughput improvements are reported for \ac{CSA} in \cite{Jakovetic15:TCOM}, while frameless ALOHA schemes are extended to the $2$- and $\nrx$-receiver case in \cite{Ogata15:FramelessMulRec} and \cite{Ogata17:CoopFrameless}, respectively. The impact of directive antennas has also been investigated in \cite{Mastilovic17:CoopSADirAntenna}, employing stochastic geometry tools to explore the applicability of multiple-receiver \ac{RA} to millimeter-wave communications.

Thanks to these contributions, a good level of maturity has been reached in understanding the gains achievable when more collectors can \emph{cooperatively} process data sent over a \ac{RA} channel. Albeit of certain interest for cellular networks, such findings do not apply to many practical \ac{MTC} systems, where \--- due to complexity or scalability reasons \--- receivers may have limited computational capabilities, and backhauling cost or bandwidth scarcity may render cooperation among them unfeasible.
In these settings, spatial diversity can still be leveraged envisioning receivers act as relays, forwarding towards a central collecting unit (part of) what decoded from users' transmissions based on the available resources. From this standpoint, a clear understanding of how much information receivers can deliver operating independently, as well as the definition of efficient strategies to accomplish the task are paramount yet still open questions for proper system design.

Starting from these remarks, we tackle in this paper a setup where users transmit following a slotted ALOHA policy towards a number of non-cooperative receivers (\emph{uplink phase}), which in turn relay packets towards a central sink (\emph{downlink phase}) over a finite-bandwidth, \ac{TDMA} channel. No \ac{SIC} capabilities are assumed, and, based on the amount of resources available in the downlink, receivers may independently decide to forward a subset of the packets they decode \--- or possibly linear combinations thereof. Besides its simplicity, the considered configuration is especially appealing in view of its scalability, as additional relays can be easily added to the system, increasing spatial diversity without the need to change the operating conditions of the network.
In this setting, the main and novel contributions we present can be summarised as follows:
\begin{itemize}
\item following an on-off fading channel model \cite{OnOff2003}, we derive exact expressions for both uplink throughput and packet loss rate for an arbitrary number of receivers;
\item assuming no information exchange among relays, we provide a lower bound on the minimum amount of downlink resources needed to asymptotically deliver to the sink with vanishingly small loss probability all information collected over the uplink. The result represents a valuable system dimensioning tool, and is especially insightful in clarifying the role played by uplink channel conditions (in terms of load and erasure rates) and by the number of available receivers;
\item the derived bound is proven to be achievable following a forwarding strategy based on random linear coding, as long as no delay constraints are set;
\item to shed light on the potential of spatial diversity when relays have limited capabilities, we propose a family of simplified forwarding strategies that require no random linear coding to be performed. Specifically, we consider the case in which each receiver independently decides upon decoding a data packet whether to store it for subsequent forwarding or to drop it. We analytically model the performance of such schemes, and optimise them based on the amount of resources granted in the downlink, possibly using simple forms of uplink channel state information locally available at the receiver;
\item the behaviour of both random linear coding schemes and simplified forwarding policies is studied when relays are equipped with a finite buffer, i.e. when constraints in terms of latency are set. Non-trivial tradeoffs emerge, revealing how simpler strategies can in fact provide competitive performance in a wide range of configurations of practical interest.
\end{itemize}

We start our discussion in Sec.~\ref{sec:sysModel} by introducing the system model, followed in Sec.~\ref{sec:uplink} by the study of uplink performance. Sec.~\ref{sec:downlinkLimits} derives the minimum downlink rates necessary for complete information retrieval at the sink and shows how this bound can be achieved via random linear coding. In Sec.~\ref{sec:downlinkSimplified} we move to the analysis of simplified forwarding strategies for the downlink, whereas Sec.~\ref{sec:buffer_downlink} compares the performance of the different donwlink strategies under delay constraints. Finally, Sec.~\ref{sec:conclusions} draws some concluding remarks.

\section{System Model and Preliminaries} \label{sec:sysModel}

\begin{figure}
\centering
\includegraphics[width=0.6\columnwidth]{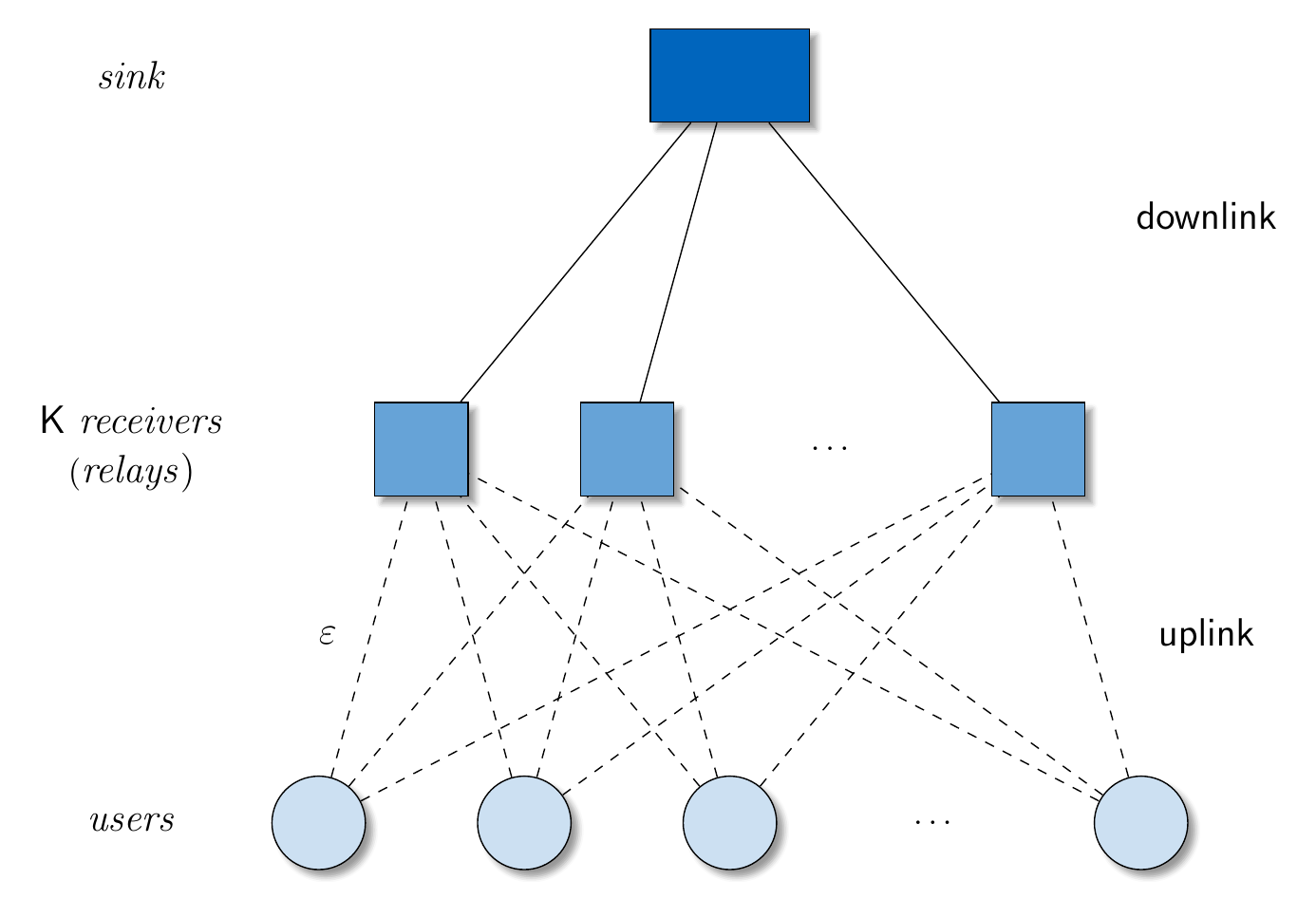}
\caption{Reference system topology: a set $\rxSet$ of \nrx\ relays collect users data over the uplink channel, later to forward them towards a sink on the downlink. Each user-relay link follows an on-off fading channel model with erasure probability~$\peras$.}
\label{fig:topology}
\end{figure}

Throughout this paper, we focus on the topology depicted in Fig.~\ref{fig:topology}, where an infinite population of users want to deliver information in the form of data packets to a collecting sink. No direct connection between users and sink is available, so that the transmission process is divided in two phases, referred to as \emph{uplink} and \emph{downlink}, respectively. During the former, data are sent in an uncoordinated fashion over a shared wireless channel to a set $\rxSet$ of $\nrx$ receivers or relays, which, in turn, forward collected information to the sink in the downlink.

As to the uplink, time is divided in successive slots of $\ulslot$ seconds, and physical layer parameters are set such that one packet can be sent within one slot. Users are assumed to be synchronised, and share the common wireless resources following a \acl{SA} \cite{Abramson:ALOHA} policy without retransmissions. Accordingly, we define the load \load\ as the average number of packets transmitted per slot, and model the number of users accessing the channel in a generic slot as a Poisson r.v. $\usersRV$ of intensity $\load$, so that $\pr\{ \usersRV = \users \} = \load^\users e^{-\load}/(\users!)$.

In order to capture the diverse propagation effects that data sent over the uplink may undergo to reach distinct receivers, we describe the wireless link connecting user $\usInd$ and relay $\rx$ as a packet erasure channel. More specifically, following the on-off fading channel introduced in \cite{OnOff2003}, we assume that a data unit is either completely shadowed  with probability $\peras$ -- not bringing any interference contribution at a receiver -- or it arrives unfaded with probability $1-\peras$. Independent realisations for any $(\usInd,\rx)$ pair as well as for a specific user-receiver couple across time slots are considered. Thanks to its simplicity, such a model allows to derive insightful closed form expressions for the key performance metrics. On the other hand, it embeds the effects of short-term receiver unavailability due, for instance, to fading or obstacles, and has been shown to effectively identify the key tradeoffs of the system under more realistic channel conditions \cite{Munari15:Massap}.

No capture effect nor multi-user detection capabilities are considered at a receiver, so that the superposition of two or more data units prevents decoding of any of them (destructive collisions). Under these assumptions, the number of non-erased packets that arrive at a relay when $\users$ concurrent transmissions take place over one slot follows a binomial distribution of parameters $(\users,1-\peras)$. Therefore, a successful reception occurs with probability $\users(1-\peras)\peras^{\users-1}$, and the average throughput experienced at each of the $\nrx$ receivers, in terms of decoded packets per slot, can be computed as:
\begin{equation}
\tpSA = \sum\limits_{\users=0}^{\infty} \frac{\load^\users e^{-\load}}{\users!} \, \users(1-\peras)\peras^{\users-1} = \load (1-\peras) e^{-\load(1-\peras)}
\label{eq:Tsa}
\end{equation}
corresponding to the performance of a SA system with erasures.
On the other hand, a diversity gain can be triggered when the  relays are considered jointly, since independent channel realisations may lead them to retrieve different information units over the same time slot. In order to quantify this beneficial effect, we label a packet as \emph{collected} when it has been received by at least one of the relays, and we introduce the \emph{uplink throughput} $\tpUL$ as the average number of collected packets per slot. Despite its simplicity, such a definition offers an effective characterisation of the beneficial effects of diversity. Indeed, it accounts for the possibility of retrieving up to $\min\{\users,\nrx\}$ distinct packets  over a slot as well as for the possibility of decoding multiple times the same data unit, as will be discussed in details in Sec.~\ref{sec:uplink}.

The \ac{RA} channel is complemented by a set of \emph{downlink} connections between the receivers and the sink, which is the final destination for information units sent by the users. In this perspective, we aim to characterise the amount of downlink resources needed to efficiently retrieve data collected over the uplink, under the constraint that no information exchange is possible among the relays. Such a problem is pivotal for system dimensioning, yet in general non-trivial due to the impossibility for a receiver to know what its peers decoded. In order to derive clear insights, we assume the downlink channel to be orthogonal to its uplink counterpart, and shared among relays by means of TDMA. Time is then divided in slots of \dlslot\ seconds each, whose duration allows to send exactly one data packet  towards the sink.
Every transmission opportunity is uniquely assigned to one relay, and slot allocation is known a priori to all receivers. Downlink channels are error-free, and relays can operate in out-of-band full-duplex mode, receiving data from the user population over the uplink while transmitting towards the sink. We restrict our attention to strategies in which each receiver re-encodes and transmits only packets it has correctly retrieved during the uplink phase, or possibly linear combinations thereof. From this standpoint, no feedback is provided by the sink, so that a relay does not know which packets have been sent by its companions and cannot adapt its forwarding choices accordingly. Finally, we initially assume infinite buffers to be available at receivers, later to relax the assumption in Sec.~\ref{sec:buffer_downlink}, where an in-depth discussion of the impact of finite queue length on the overall system performance is presented.

\section{Uplink Performance}  \label{sec:uplink}

To gauge the impact of receiver diversity, we start focusing on the throughput achievable over the \ac{SA}-operated uplink channel. Leaning on the definition introduced in Sec.~\ref{sec:sysModel}, let $\collRV$ be the r.v. with alphabet $\{0,1,2, \ldots, \nrx\}$ that describes the number of packets collected by the relays over one slot.
\begin{prop}
The uplink throughput of the considered multi-receiver \ac{SA} channel with erasures evaluates, for any $\nrx >1$, to
 \begin{equation}
 \tpUL = \nrx \,\tpSA -\sum_{\rx=2}^\nrx (-1)^{\rx} {\nrx \choose \rx} \load (1-\peras)^\rx e^{-\load(1-\peras^\rx)} \,.
 \label{eq:truUplink}
\end{equation}
\label{prop:ulTru}
\end{prop}
\vspace{2mm}
\begin{IEEEproof}
Under the assumptions of Sec.~\ref{sec:sysModel}, the number of packets collected by the set of relays within a slot, captured by the r.v. $\collRV$, is independent and identically distributed over different time units. By its very definition, moreover, \mbox{$\tpUL = \mathbb E[ \collRV]$}.  Let us now observe the behaviour of the uplink channel for $\numulslot$ subsequent slots, and let $\setArx$ be the set of data units decoded by relay $\rx$ during the whole interval. Accordingly, the overall set of collected packets can be expressed as $\bigcup_{ \rx \in \rxSet } \setArx$. By the weak law of large numbers, we have
\begin{equation}
\tpUL = \lim_{\numulslot\rightarrow\infty} \frac{1}{\numulslot} \left| \,\bigcup\nolimits_{\rx=1}^\nrx \setArx \,\right|
\label{eq:proofTruK_tpUL}
\end{equation}
or, more formally
\begin{equation}
\lim_{\numulslot\rightarrow\infty} \pr\left\{ \,\left\lVert \,\,\frac{1}{\numulslot} \cdot \left| \,\bigcup\nolimits_{\rx=1}^\nrx \setArx \,\right| - \tpUL \,\,\right\rVert  \geq \sv\,\right\} = 0, \hspace*{.7em} \forall \sv \in \mathbb R^+
\end{equation}
where $\lVert \cdot \rVert$ indicates the absolute value.
Let us now denote by $\rxSubSet \subseteq \rxSet$ a subset of relays of cardinality $|\rxSubSet| = \nrxSubSet$. By virtue of the  inclusion-exclusion principle (see, e.g. \cite{slomson1991introduction}), we have
\begin{equation}
\left| \, \bigcup\nolimits_{\rx=1}^\nrx \setArx \, \right| = \sum\limits_{\rxSubSet\subseteq \rxSet,\,\rxSubSet \not = \emptyset} (-1)^{\nrxSubSet-1} \left| \bigcap\nolimits_{ \subsetrx \in \rxSubSet } \setAsubset \right| .
\label{eq:proofTruK_inclExcl}
\end{equation}
The derivation of \tpUL, thus, simply requires to compute the cardinality of $\bigcap_{ \subsetrx \in \rxSubSet } \setAsubset$ for asymptotically long uplink observation intervals (i.e. $\numulslot \rightarrow \infty$). Furthermore, due to the symmetry of the topology, the sought value does not depend on the specific receivers being considered, and can be determined for any of the $\binom{\nrx}{\nrxSubSet}$ subsets of $\nrxSubSet$ relays. For an arbitrary set $\rxSubSet$, \eqref{eq:proofTruK_inclExcl} takes the form
\begin{equation}
\left| \, \bigcup\nolimits_{\rx=1}^\nrx \setArx \, \right| = \sum\limits_{\nrxSubSet=1}^\nrx \binom{\nrx}{\nrxSubSet} (-1)^{\nrxSubSet-1} \left| \bigcap\nolimits_{ \subsetrx \in \rxSubSet } \setAsubset  \right|.
\label{eq:proofTruK_inclExcl_simplified}
\end{equation}
At each slot, the cardinality of the set of packets decoded in common by the $\nrxSubSet$ relays can increase by at most one unit. This occurs when the same packet is retrieved by all the considered receivers, i.e. with probability
\begin{equation}
\psamecoll = \sum\limits_{\users=0}^{\infty} \frac{\load^\users \, e^{-\users}}{\users!} \, \users \left[  (1-\peras) \peras^{\users-1} \right]^\nrxSubSet
= \load \, (1-\peras)^\nrxSubSet \,e^{-\load (1-\peras^\nrxSubSet)}\,.
\label{eq:proof_pjVal}
\end{equation}

Recalling the independence of transmission patterns and erasures, $\left|\bigcap_{ \subsetrx \in \rxSubSet } \setAsubset\right|$ is then described by a geometric r.v. with parameters $(\numulslot,\psamecoll)$. Leaning on this result, we can combine \eqref{eq:proofTruK_tpUL} and \eqref{eq:proofTruK_inclExcl_simplified} to get
\begin{align}
\tpUL &= \lim_{\numulslot\rightarrow\infty} \, \frac{1}{\numulslot}  \sum\limits_{\nrxSubSet=1}^\nrx \binom{\nrx}{\nrxSubSet} (-1)^{\nrxSubSet-1} \left| \bigcap\nolimits_{ \subsetrx \in \rxSubSet } \setAsubset \right| \\
&= \sum\limits_{\nrxSubSet=1}^\nrx (-1)^{\nrxSubSet-1} \binom{\nrx}{\nrxSubSet} \,\load \, (1-\peras)^\nrxSubSet \,e^{-\load (1-\peras^\nrxSubSet)}
\end{align}
where the second equality stems by applying the law of large numbers to the r.v. $\left|\bigcap_{ \subsetrx \in \rxSubSet } \setAsubset\right|$ of expected value $\psamecoll \numulslot $. The formulation in \eqref{eq:truUplink} follows by isolating the term for $\nrxSubSet=1$.
\end{IEEEproof}
\begin{figure}
\centering
\includegraphics[width=.6\columnwidth]{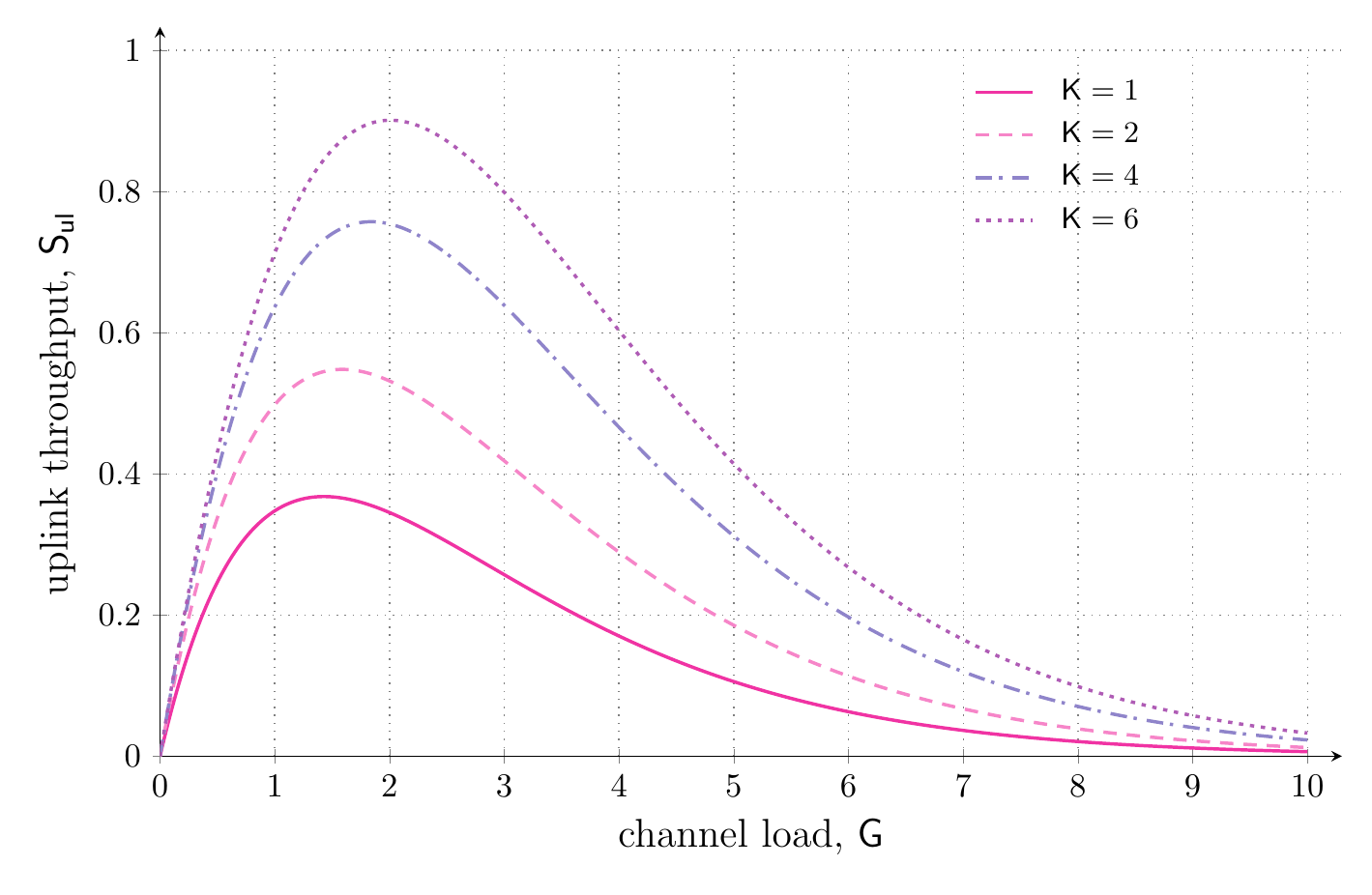}
\vspace{-.5em}
\caption{Average uplink throughput vs. channel load for different number of relays $\nrx$. The erasure probability has been set to $\peras=0.3$.}
\label{fig:truputUplink}
\end{figure}

Prop.~\ref{prop:ulTru} conveniently expresses \tpUL\ as the difference between \nrx\ times the throughput of a single receiver -- an intuitive upper bound to the uplink performance -- and a correcting term, which accounts for the possibility of having the same information unit redundantly decoded at more than one relay. The beneficial impact of receiver diversity is highlighted in  Fig.~\ref{fig:truputUplink}, which depicts \tpUL\ against the channel load for different values of \nrx\ and an erasure probability $\peras=0.3$. In the presence of a single relay (solid line), the system behaves as a plain \ac{SA} channel, with a peak efficiency of $e^{-1}\simeq 0.36$ decoded packets per slot obtained for $\load = (1-\peras)^{-1}$. Conversely, the availability of a second receiver boosts the maximum achievable throughput by $\sim 50\%$, and up to $0.9\,\rm{pkt/slot}$ are retrievable for $\nrx=6$, without any modification to the medium access policy at the transmitter side. Such a result stems from two main factors. On the one hand, broader receiver sets enable a given slot to see larger peaks of throughput, as up to \nrx\ data units can be collectively retrieved. On the other hand, the introduction of additional relays improves the decoding probability even when fewer than \nrx\ users accessed the channel, thanks to the independent erasure patterns packets experience to reach distinct receivers.

\begin{figure}
\centering
\includegraphics[width=.6\columnwidth]{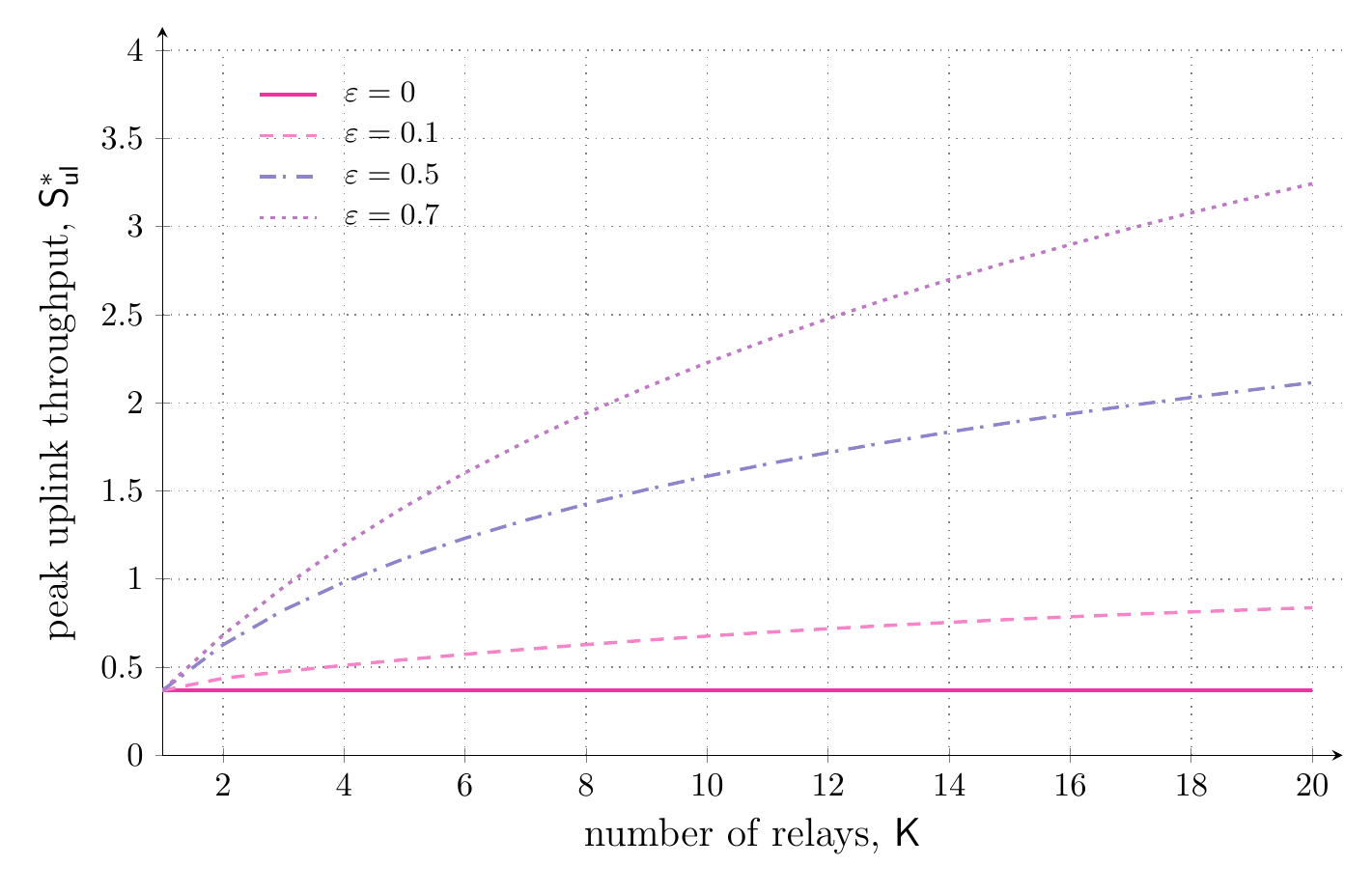}
\vspace{-.5em}
\caption{Maximum achievable uplink throughput $\tpULMax$ as a function of the number of relays $\nrx$, for different erasure rates $\peras$.}
\label{fig:peakTruUplink}
\end{figure}

The reported trends raise the natural design question of how \nrx\ shall be set to properly hit the performance-cost tradeoff between higher uplink performance and deployment of more relays. Fig.~\ref{fig:peakTruUplink} delves into this aspect by showing, for different erasure probabilities, the maximum achievable throughput $\tpULMax$ when increasing the number of receivers.\footnote{Due to the transcendental nature of the addends in \eqref{eq:truUplink}, no closed form expression for the peak throughput exists. Nonetheless, $\tpULMax$ can easily be derived via numerical maximisation techniques for any value of \nrx\ and \peras.} In ideal channel conditions ($\peras=0$), larger values of \nrx\ bring no benefit to the system, as the same set of transmitted packets reach all of the relays, calling off any diversity. Conversely, in the presence of erasures, $\tpULMax$ monotonically increases with \nrx. From this standpoint, two remarks are in order. Firstly, the plot highlights how the improvement triggered by additional receivers progressively reduces, leading to a growth rate for the achievable throughput that is less than linear in $\nrx$. Following this diminishing-returns behaviour, the most appealing advantages are reaped for rather small -- and thus practically viable -- relay sets. Secondly, from Fig.~\ref{fig:peakTruUplink} it is manifest that higher erasure rates ($\peras<1$) favour a decorrelation in the pattern of packets that can be retrieved at receivers, boosting diversity and consequently improving the achievable throughput.
Along this line, if the uplink throughput is the main goal, e.g. for \ac{MTC} applications collecting possibly redundant data from a large population of sensors, multi-receiver \ac{SA} might be a simple and effective solution when working at high load and in harsh channel conditions.

\begin{figure}
\centering
\includegraphics[width=.6\columnwidth]{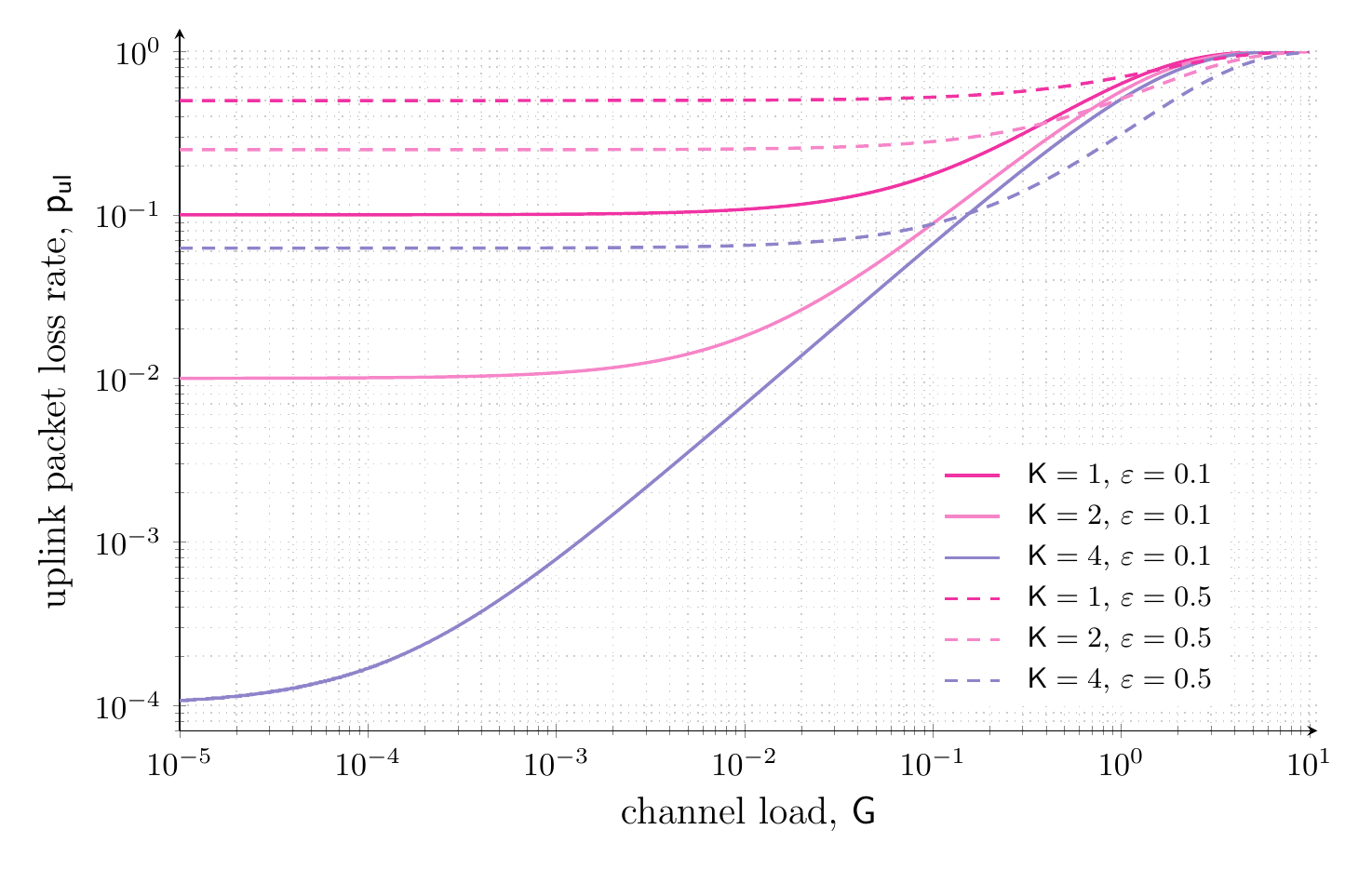}
\vspace{-.5em}
\caption{Packet loss rate $\plrUL$ vs. channel load for different erasure probabilities and cardinalities of the relay set.}
\label{fig:plr}
\end{figure}

On the other hand, higher aggregate data collection rates are obtained for progressively larger values of \load, and come at the expense of poorer per-user reliability. In the classical single-receiver case, for instance, simple probabilistic arguments reveal that a sent data unit is not retrieved with probability $1-(1-\peras) e^{-\load(1-\peras)}$. If the load is set to maximise \tpUL, i.e. $\load=(1-\peras)^{-1}$, the loss rate is thus larger than \mbox{$1-e^{-1}\simeq0.63$} regardless of \peras, a value which is hardly tolerable for a wide class of applications. In view of this, many practical wireless systems rely on \ac{SA} in lightly loaded channels,  e.g. for logon and control signalling, aiming at reliability rather than aggregate throughput.  To understand whether  diversity can improve performance in this operating conditions as well, we focus on the packet loss rate $\plrUL$. The metric is  defined as the probability that a sent data unit is not collected -- either due to fading or to collisions -- at any of the relays.

\begin{prop}
For any $\nrx \geq 1$ and $\peras$, the packet loss rate of the considered uplink channel is given by
\begin{equation}
\plrUL = \sum_{\rx=0}^\nrx (-1)^\rx {\nrx \choose \rx} (1-\peras)^\rx e^{-\load(1-\peras^\rx)} \,.
\label{eq:plr}
\end{equation}
\label{prop:plr}
\end{prop}
\begin{IEEEproof}
Let us focus on the packet transmitted by user \usInd, and let $\interfRV$ be the r.v. describing the number of data units sent on the channel over the same slot. Conditioning on $\interfRV = \interfrv$, a relay retrieves the packet of interest only if the latter arrives unfaded while all interferers are erased, i.e. with probability $(1-\peras) \peras^\interfrv$. By virtue of the independence of all user-to-relay connections, the conditional probability for the data unit not to be collected evaluates to \mbox{$[\,1- (1-\peras) \peras^\interfrv\,]^{\nrx}$}. Resorting to the binomial theorem, the expression can be conveniently reformulated as:
\begin{align}
\pr\{  \usInd \text{ not collected}  \,|\, \interfRV = \interfrv \} = \sum_{\rx=0}^\nrx (-1)^\rx {\nrx \choose \rx} \left[(1-\peras)\peras^{\mathsf i}\right]^\rx.
\end{align}
On the other hand, the number of interferers seen by a user that accesses the channel still follows a Poisson distribution of intensity $\load$. Leaning on this, the proposition statement  follows by removing the conditioning on $\interfRV$.
\end{IEEEproof}

The compact expression in \eqref{eq:plr} provides useful design guidelines, quantifying the maximum \load\ that can be supported on the uplink channel while guaranteeing a desired packet loss rate. More insights are provided by Fig.~\ref{fig:plr}, which reports $\plrUL$ against the channel load. Different colours indicate the behaviour when $1$, $2$ and $4$ receivers are available, with solid lines referring to an erasure rate $\peras=0.1$ and dashed ones to the $\peras=0.5$ case. As expected, good reliability levels from the single-user perspective are only achieved for low channel occupations. Remarkably, the availability of multiple relays triggers a dramatic improvement in this operating region, enabling delivery rates that would otherwise not be possible irrespective of the channel load. Indeed, when \mbox{$\load\rightarrow 0$}, failures are mainly induced by fading events, and $\plrUL$ approaches $\peras^\nrx$, so that an additional receiver can lower the packet loss probability by up to a factor $\peras$. The simple asymptotic expression for $\plrUL$ also pinpoints the detrimental role of higher erasure rates on the success rate experienced by a user in a lightly loaded system, well exemplified in the plot comparing the trends for $\peras=0.1$ and $\peras=0.5$ . When the channel gets more congested, instead, harsher $\peras$ values end up having a beneficial influence. In such conditions, in fact, higher erasure rates positively reduce the cardinality of the set of overlapping packets, potentially solving collisions and enabling retrieval of one data unit. The effect becomes dominating for large enough \load, inducing the intersection of curves reported in Fig.~\ref{fig:plr} and buttressing the throughput trends discussed in Fig.~\ref{fig:peakTruUplink}.

\section{Downlink Performance Limits}  \label{sec:downlinkLimits}

Leaning on the uplink characterisation, we now tackle the task of efficiently delivering collected packets to the sink, under the constraint that no information exchange is allowed among receivers. We recall that relays share the error-free yet finite-bandwidth downlink resources via \ac{TDMA}. Furthermore, we assume operations to be split in successive phases of duration $\slotTime$, each of which sees $\numulslot$ and $\numdlslot$ slots elapse in the uplink and downlink, respectively.\footnote{In other words, system parameters are set so that \mbox{$\numulslot \ulslot = \numdlslot \dlslot = \slotTime$.}} Accordingly, the \mbox{$\rx$-th} relay observes the \ac{SA} channel for $\slotTime$ seconds, and then independently decides what to forward to the sink during the next phase over the slots it was assigned.
Following this notation, we define the downlink sum-rate $\dlrate$ as the number of downlink transmissions allocated per uplink slot, given by $\dlrate := \ulslot/\dlslot = \numdlslot/\numulslot$. In the same way, the downlink rate of relay $\rx$ follows as $\dlrate_\rx := \dltxk \dlrate$, where $\dltxk \in[0,1]$ indicates the fraction of the \numdlslot\ transmission opportunities allotted to $\rx$, and \mbox{$\sum_{\rx=1}^{\nrx} \dltxk = 1$}.

Within this framework, a pivotal aspect of proper design lies in how to dimension the receiver-to-sink channel. As a first step, we focus in this section on the asymptotic performance of  the system, and allow an arbitrarily long observation period of the uplink (i.e. $\numulslot \rightarrow \infty$). Under this assumption, we aim at determining rates that enable the sink to retrieve the whole information content gathered by the set of receivers with high probability. A straightforward solution follows recalling that relays individually enjoy the average throughput of a \ac{SA} channel. Thus, providing $\tpSA$ transmission opportunities per uplink slot to each of them (i.e. $\dlrate = \nrx \tpSA$) indeed grants the sink to asymptotically get all the packets that were received. Such an allocation, however, is in general not efficient, since a data unit decoded by more than one relay will be delivered over the error-free downlink multiple times, entailing a waste of bandwidth. We are instead interested in determining the minimum downlink sum rate \dlratemin\ to accomplish the task or, more formally, in
\begin{equation}
\dlratemin := \min\left\{ \dlrate \, | \, \plos \rightarrow 0, \numulslot \rightarrow \infty \right\}
\end{equation}
where $\plos$ is the probability that a packet collected over the uplink cannot be decoded by the sink due to the downlink forwarding policy.

To characterise $\dlratemin$ we follow a two-step approach, first deriving a lower bound (Sec.~\ref{sec:bounds}) and then  proving its achievability via a simple forwarding strategy based on \ac{RLC} (Sec.~\ref{sec:nc}).

\subsection{A lower bound on the minimum downlink sum-rate}
\label{sec:bounds}

The minimum amount of downlink resources needed for reliable data delivery to the sink can be derived leaning again on combinatorial techniques.

\begin{prop}
For any $\rxSubSet \subseteq \rxSet$ with cardinality $|\rxSubSet|=\nrxSubSet$, a set of rates $\{ \dlrate_\rx \, | \, \rx \in \rxSubSet, \plos \rightarrow 0 \}$ have to satisfy
\begin{equation}
\sum_{\rx \in \rxSubSet} \dlrate_\rx \geq \tpUL - \omega\left( \nrxSubSet, \load, \peras \right)
\label{eq:rateBoundSet}
 \end{equation}
 where the ancillary function $\omega$ is defined as
\begin{equation}
 \omega \left( \nrxSubSet, \load, \peras \right) := \sum\limits_{\ell=1}^{\nrx - \nrxSubSet} (-1)^{\ell-1} { \nrx - \nrxSubSet \choose \ell } \load (1-\peras)^\ell e^{-\load (1-\peras^\ell)}.
 \end{equation}
 \label{prop:dlRatesBound}
\end{prop}
\vspace{1mm}
\begin{IEEEproof}
The argument follows the notation and approach introduced in the proof of Prop.~\ref{prop:ulTru}. Let us then consider a subset  of relays $\rxSubSet \subseteq \{1,\dots,\nrx\}$, and indicate by $\bigcup_{ \rx \in \rxSubSet } \setArx$ the set of packets collected by them over the $\numulslot$ observed slots. In order to grant successful recovery at the sink, at least all data units that have been retrieved only by nodes in $\rxSubSet$ (and not by anyone else) have to be forwarded in the downlink. By the definition of downlink rates, this translates into
\begin{equation}
\sum_{\rx \in \rxSubSet} \numulslot \, \dlrate_\rx \geq \left| \, \bigcup\nolimits_{\rx\in \rxSubSet} \setArx \,\backslash\, \bigcup\nolimits_{\rx \in \overline{\rxSubSet}} \,\setArx \,\right|
\label{eq:ratebound1}
\end{equation}
where $\overline{\rxSubSet} := \rxSet \,\backslash\, \rxSubSet$. If we now observe that
\begin{equation}
\bigcup\nolimits_{\rx\in \rxSubSet} \setArx \,\backslash\, \bigcup\nolimits_{\rx \in \overline{\rxSubSet}} \,\setArx = \bigcup\nolimits_{\rx\in \mc \rxSet} \setArx \,\backslash\, \bigcup\nolimits_{\rx \in \overline{\rxSubSet}} \,\setArx
\end{equation}
the inclusion-exclusion principle can be applied in its complementary form, to obtain
\begin{align}
&\left| \, \bigcup\nolimits_{\rx\in \rxSubSet} \setArx \,\backslash\, \bigcup\nolimits_{\rx \in \overline{\rxSubSet}} \,\setArx \,\right| = \\[1mm]
&\hspace*{5mm}\left| \bigcup\nolimits_{\rx\in \mc \rxSet} \setArx \right| - \!\!\sum \limits_{\dummy=1}^{\nrx-|\overline{\rxSubSet}|} (-1)^{\dummy-1} {\nrx - |\overline{\rxSubSet} |  \choose \dummy}\left| \bigcap\nolimits_{ \subsetrx \in \rxSubSet } \setAsubset \right| .\,\,\,\,\,\,\,
\label{eq:proofCardinality}
\end{align}
As discussed within Prop.~\ref{prop:ulTru} for the simplification of \eqref{eq:proofTruK_inclExcl}, the cardinality of $\bigcap_{ \subsetrx \in \rxSubSet } \setAsubset$ follows a geometric distribution with parameters $(\numulslot,\psamecoll)$, with $\psamecoll$ given in \eqref{eq:proof_pjVal}. Plugging \eqref{eq:proofCardinality} into \eqref{eq:ratebound1} and applying the law of large numbers   leads to the statement formulation. Indeed, for $\numulslot \rightarrow \infty$, the first addend in \eqref{eq:proofCardinality} -- expressing the number of collected packets by the relay set -- converges to $\numulslot \tpUL$, whereas the $\dummy$-th term within the summation tends to $(-1)^{\dummy-1} \numulslot \,\psamecoll$.
\end{IEEEproof}

An application of the condition to the whole set of receivers (i.e. for $\rxSubSet = \rxSet$) provides a compact characterisation of $\dlratemin$, summarised by Lemma~\ref{lemma:sumRate}.
\begin{lemma}
The minimum downlink sum-rate is bounded from below as $\dlrate^* \geq \tpUL $.
\label{lemma:sumRate}
\end{lemma}

The result confirms the basic intuition that we need to grant in the downlink at least as many transmissions on average as the number of collected packets per slot in the uplink. Moreover, combined with \eqref{eq:truUplink}, it provides for any $(\load, \peras, \nrx)$ tuple a simple system dimensioning tool. On the other hand, approaching the bound might not be trivial in general, as no form of information exchange among relays on the data set each of them has collected is permitted.

\subsection{Achievability via \acl{RLC}} \label{sec:nc}

To prove that the bound derived in Sec.~\ref{sec:bounds} is indeed achievable, we consider a forwarding strategy based on \ac{RLC}. To this aim, let us focus on relay $\rx$. Over any of the \numulslot\ uplink slots, $\rx$ either decodes the $\payload$-bit payload sent by a user or does not retrieve any information (due to a destructive collision, to erasures of all packets, or to the absence of any active transmitter). We thus describe the set of observations at the receiver via the $(\numulslot\times 1)$ vector $\linComb_\rx$ of elements in $\mathbb F_{\fieldOrd} \cup \,\erasure$, where $\mathbb F_{\fieldOrd}$ is a finite extension field of order $\fieldOrd = 2^\payload$ and $e$ accounts for the event of no decoding. Transmission opportunities in the downlink are in turn used by $\rx$ to send linear combinations of what collected. Specifically, the relay generates an $(\numulslot \dlratek \times \numulslot)$ coefficient matrix $\MxCombK{\rx}$, whose $\slot$-th column is \mbox{all-zero} if $\erasure$ was observed over slot $\slot$, and composed of elements drawn uniformly at random from $\mathbb F_\fieldOrd$ otherwise. The $\numulslot\dlrate_\rx$ packets to be forwarded are then obtained as $\colVecK{\rx} = \MxCombK{\rx} \linCombK{\rx}$.\footnote{Coefficients used for generating linear combinations are also available at the sink for decoding. This can be achieved by having the relay piggyback them in the packet header (adding overhead), or by generating the matrix $\MxCombK{\rx}$ according to some previously agreed pseudo-random sequence know both at the receiver and the sink.}
Combinations sent by all receivers are correctly delivered over the error-free downlink connections, leading to a system of $\numulslot \dlrate$ equations in the form:
\begin{equation}
{\underbrace{ \left(
\begin{array}{c}
\colVecK{1} \\ \colVecK{2} \\ \vdots \\ \colVecK{\nrx}
\end{array}
 \right) }_{\colVec}} =
 \underbrace{\left(\begin{array}{cccc} \MxCombK{1} & 0 & \ldots & 0\\0 & \MxCombK{2} & \ldots & 0 \\ 0 & 0 & \ddots & 0 \\ 0 & 0 & \ldots & \MxCombK{\nrx} \end{array}\right)}_{\MxComb}
\underbrace{\left( \begin{array}{c} \linCombK{1}\\ \linCombK{2} \\ \vdots \\ \linCombK{\nrx} \end{array} \right)}_{\linComb}
\label{eq:sysComplete}
\end{equation}
where $\colVec$ and $\linComb$ are $(\numulslot \dlrate \times 1)$ and $(\numulslot \nrx \times 1)$ column vectors, respectively, while $\MxComb$ is a $(\numulslot  \dlrate \times \numulslot \nrx)$ matrix. Solving \eqref{eq:sysComplete}, packets originally transmitted by the user population can eventually be gathered at the sink. Remarkably, this simple and uncoordinated approach makes it possible to deliver the whole uplink throughput using the minimum amount of downlink resources, as we prove below:
\begin{prop}
For any set of downlink rates \mbox{$\{ \dlrate_\rx \,|\, \rx\in \rxSet \}$} satisfying \eqref{eq:rateBoundSet}, the discussed \ac{RLC} forwarding strategy enables retrieval at the sink of all packets collected by the set of receivers with high probability as $\numulslot \rightarrow \infty$.
\label{prop:rlcAchiev}
\end{prop}
\begin{IEEEproof}
We start by observing that the system of equation in \eqref{eq:sysComplete} may contain redundant components.  Indeed, all elements of $\linComb$ equal to $\erasure$ do not contribute to information retrieval, reporting the outcome of slots without successful decoding. Moreover, due to the lack of communication among relays, some elements of $\linComb$ may be identical, identifying packets decoded by more than one receiver. Let us therefore indicate by  \mbox{$\colVec = \MxCombRed \linCombRed$} a reduced version of the original system, where columns of from $\MxComb$ are removed (if corresponding to erasures) or linearly combined (if  corresponding to duplicate packets), and $\linCombRed$ contains only the $|\bigcup\nolimits_{\rx=1}^{\nrx} \setArx|$ distinct received packets and no $\erasure$ symbols.

\begin{figure*}[!bp]
\normalsize
\hrulefill
\begin{align}
 \left( \begin{array}{c} \colVecK{1}\\ \colVecK{2} \end{array} \right) =
  \left(\begin{array}{c}\MxCombRedK{1}\\ \MxCombRedK{2}
                            \end{array}\right)
                            \linCombRed
 =
 \left(\begin{array}{ccc} \MxCombRedK{1,\{1\}} & \bm 0 & \MxCombRedK{1,\{1,2\}} \\
                \bm 0 & \MxCombRedK{2,\{2\}} & \MxCombRedK{2,\{1,2\}}
                            \end{array}\right)
 \left( \begin{array}{l} \linCombRedK{\{1\}}\\ \linCombRedK{\{2\}} \\ \linCombRedK{\{1,2\}}
 \end{array} \right)
 \label{eq:exampleG_K=2}
\end{align}
\end{figure*}

The entries in $\linCombRed$ can be partitioned into the $2^\nrx-1$ vectors $\linCombRedK{\rxSubSet}$, one for every non-empty set $\rxSubSet\subseteq \rxSet$. Each $\linCombRedK{\rxSubSet}$ contains all packets that have been received \emph{only} by relays in $\rxSubSet$, i.e. the set \mbox{$\USet_{\rxSubSet} := \bigcap_{\rx\in \rxSubSet} \setArx \backslash \bigcup_{\rx \in \overline{\rxSubSet}} \setArx$}. Leaning on this, the columns in $\MxCombRed$ and the rows in $\linCombRed$ can be permuted to obtain
\begin{align}
 \colVecK{\rx} = \MxCombRedK{\rx} \linCombRed:= \sum_{\rxSubSet\subseteq \rxSet} \MxCombRedK{\rx,\rxSubSet} \,\linCombRedK{\rxSubSet}, \quad \forall~\rx=1,\ldots,\nrx 
\end{align}
where the $(\numulslot \dlrate_\rx \times |\,\USet_{\rxSubSet}\,|)$ submatrix $\MxCombRedK{\rx,\rxSubSet}$ contains only elements from $\mathbb F_{\fieldOrd}$ if $\rx \in \rxSubSet$ and is all-zero otherwise. An example of the simplified system structure for the $\nrx=2$ case is shown in \eqref{eq:exampleG_K=2} at the bottom of the page.

Let us now focus on the downlink transmissions performed by relays in a subset $\rxSubSet$. The variables involved only in the corresponding $\numulslot \sum\nolimits_{\rx\in\rxSubSet} \dlrate_\rx$ equations are those grouped in the vectors $\linCombRed_{\LSet}$, $\forall \LSet \subseteq \rxSubSet$. A necessary condition for decoding is that the number of equations be larger than or equal to the number of unknowns, i.e.
\begin{equation}
 \numulslot \sum_{\rx \in \rxSubSet}\dlrate_i \geq \sum_{\LSet \subseteq \rxSubSet}  \left| \,\USet_{\LSet}\right| =  \left| \bigcup\nolimits_{\rx\in \rxSubSet} \setArx \,\backslash \, \bigcup\nolimits_{\rx \in \overline{\rxSubSet}} \setArx \right|
 \end{equation}
where the rightmost equality is proven in App.~\ref{app:achievability}. Recalling the proof of Prop.~\ref{prop:dlRatesBound}, the requirement is met for any rate allocation satisfying \eqref{eq:rateBoundSet} when $\numulslot \rightarrow \infty$. Indicate now as $\MxCombRed_{\rxSubSet}$ the ($\numulslot \sum\nolimits_{\rx\in\rxSubSet} \dlrate_\rx$)-row submatrix of $\MxCombRed$ obtained considering only $\MxCombRed_{\rx}$, $\rx\in\rxSubSet$. A sufficient condition to retrieve all unknowns in $\linCombRed_{\rxSubSet}$ is for $\MxCombRed_{\rxSubSet}$ to have rank $\sum_{\mc L \subseteq \rxSubSet}  \left| \,\mc U_{\mc L}\right|$. Denote the set of indices of non-zero columns of matrix $\MxCombRedK{\rx}$ as the support of $\MxCombRedK{\rx}$. By construction, a row of $\MxCombRedK{\rx}$ has a different support than a row of $\MxCombRedK{j}$, for $\rx \not = j$. These rows are thus linearly independent. On the other hand, since all nonzero elements are randomly drawn from $\mathbb F_{\fieldOrd}$, the probability of linear dependence among rows of a submatrix $\MxCombRedK{\rx}$ can be made arbitrarily small by picking a large enough $\fieldOrd$ \cite{Lidl96_FiniteFields}, granting the sufficient rank.
The packets collected over the uplink by any subset of relays $\rxSubSet$ can then be retrieved at the sink. Applying the result for $\rxSubSet = \rxSet$ proves the statement.
\end{IEEEproof}
\section{Downlink Performance: Analysis of Some\\ Practical Forwarding Policies} \label{sec:downlinkSimplified}

As clarified in Sec.~\ref{sec:downlinkLimits}, a \ac{RLC} forwarding strategy can deliver to the sink with high probability all information collected over the uplink resorting to the minimum amount of resources. When brought to implementation, however, this approach incurs drawbacks that partly counterbalance its benefits. In fact, an increased complexity is triggered both at the relays and at the sink to generate and process linear combinations of data units. Moreover, an efficiency cost in terms of bandwidth arises if coefficients employed to encode the transmitted data units are piggybacked onto packet headers. Both aspects may become critical when the uplink is observed over a long time interval prior to triggering downlink transmissions, a necessary condition for \ac{RLC} to be effective. From this standpoint, the definition of simpler strategies becomes relevant to unleash the potential of receiver diversity in practical settings.

Let us then assume that no packet-level coding across collected data units is possible.  Recalling that the uplink channel is fed via \ac{SA}, as soon as $\dlrate_\rx < \tpSA$, the downlink resources available to receiver $\rx$ are not sufficient on average to forward all packets it retrieves. The relay has then to selectively decide which units to place on the downlink channel. This condition is epitomised by a policy in which, upon decoding a packet, the receiver may either drop it or enqueue it for later transmission. In general, the decision can be made considering side-information on the state of the uplink channel, leading to the following definition which will serve as reference for our discussion:

\emph{Definition: \ac{SFP}}. Let $\{\mc E_j\}$ be a $(\ms Q +1)$-element partition of $\mathbb N$. A \ac{SFP} associates to each relay $\rx \in \rxSet$ a vector $\dropVec_{\rx} = [\dropij{\rx}{1}, \dots, \dropij{\rx}{\ms Q+1}]$ with elements in the interval $[0,1]$, representing enqueueing probabilities. For an uplink slot of interest, denote by $\users \in \mc E_j$ the number of packets transmitted by the users. If $\rx$ decodes one of them, it discards the packet with probability $1-\dropij{\rx}{j}$, or enqueues it in a FIFO buffer with probability $\dropij{\rx}{j}$ for subsequent downlink transmission.

Based on this definition, receiver $\rx$ can be modelled as an infinite queue with average arrival rate
\begin{equation}
\arrRate{\rx} := \sum_{j=1}^{\ms Q+1} \sum_{\users \in \mc E_j} \frac{\load^\users e^{-\load}}{\users!} \, \users \,(1-\peras) \peras^{\users-1} \cdot \dropij{\rx}{j}
\label{eq:bufferArrivalRate}
\end{equation}
with $\arrRate{\rx} = \tpSA$ when no packet is dropped, i.e. for \mbox{$\dropij{\rx}{j} = 1$}, $\forall j$. We then restrict our study to downlink dimensionings that ensure all enqueued data units to be delivered to the sink, considering rate allocations in the form $\dlrate_\rx = \arrRate{\rx}$.\footnote{Strictly speaking, buffers are stable for any rate $\dlrate_\rx > \arrRate{\rx}$. Results shall then be interpreted as a downlink dimensioning $\dlrate_\rx = \arrRate{\rx} + \delta$, $\delta \rightarrow 0$.}

As opposed to \ac{RLC}-based schemes, such strategies do not entail any complexity in terms of packet-level coding, offering a viable alternative for the downlink. However, they are inherently not able to ensure delivery of all decoded data units as soon as some $\dropij{\rx}{j} < 1$, since a packet may be discarded by every relay that retrieved it.
To further investigate this tradeoff we consider and study two classes of \ac{SFP} for the $\nrx=2$ case, optimising the probability vectors $\dropVec_\rx$ so as to maximise information delivered to the sink for a given sum-rate $\dlrate$. Performance is evaluated by means of the downlink throughput $\tpDL$, defined as the average number of user-generated packets that are retrieved at the relays and eventually reach the sink per uplink slot. Moreover, to gauge the effectiveness of different forwarding policies, we compare their performance to the one of a reference benchmark, given by the maximum achievable downlink throughput for a given rate allocation
\begin{equation}
\capDL(\dlrate) = \sup\{ \,\tpDL \,|\, \dlrate, \plos \rightarrow 0\,\}
\end{equation}
where $\plos$ indicates the probability for a packet collected over the uplink \emph{and enqueued} at relays not to be retrieved at the sink. Leaning on the results of Sec.~\ref{sec:downlinkLimits}, we get:
\begin{coroll}
The maximum achievable downlink throughput of the system for $\nrx=2$ satisfies:
\begin{equation}
\capDL(\dlrate) = \left\{
\begin{aligned}
\dlrate & \quad \mbox{for } \dlrate < \tpUL \\
\tpUL & \quad \mbox{for } \dlrate \geq \tpUL \\
\end{aligned}
\right.
\label{eq:Cdl}
\end{equation}
\label{th:capRegion}
\end{coroll}
\begin{IEEEproof}
For $\dlrate \geq \tpUL$, the result is simply a reformulation of Prop.~\ref{prop:dlRatesBound} and \ref{prop:rlcAchiev} when $\nrx=2$. Conversely, let $\beta=\dlrate/\tpUL<1$, and assume that each relay drops a packet received over the uplink with probability $1-\beta$. It follows that the average number of collected data units evaluates to $\beta\, \tpUL$, so that the downlink phase is equivalent to the one of a system serving an uplink throughput of $\dlrate$ packets per slot. The propositions of Sec.~\ref{sec:downlinkLimits} apply to the scaled downlink, proving the result.
\end{IEEEproof}

\subsection{Uplink-channel agnostic simplified forwarding policies}
We first consider the simplest case in which relays have no side-information on the uplink channel state, and buffering decisions are made irrespective of the number of packets that were transmitted over the slot. Based on the definition of a \ac{SFP}, the setup corresponds to $\ms Q=0$, \mbox{$\{ \mc E_j\} = \mc E_1 = \mathbb N$}, and the downlink phase of this \emph{uplink-agnostic} strategy is completely specified by the pair $(\drop_1,\,\drop_2)$.\footnote{We omit the subscripts in $\dropij{1}{1}$ and $\dropij{2}{1}$ for the sake of readability.}  The average number of transmissions in the downlink follows from \eqref{eq:bufferArrivalRate} as $\dlrate = (\drop_1 + \drop_2)\,\tpSA$. Moreover, since all buffered data units are delivered to the sink, $\tpDL$ can be computed as the average number of \emph{distinct} packets (i.e. counting duplicates only once) that are enqueued by the set of relays per uplink slot. Simple combinatorial arguments reveal that the probability of having the same information unit enqueued by both receivers conditioned on having $u$ users transmitting is given by $\drop_1 \drop_2 u (1-\peras)^2 \peras^{2(u-1)}$. Averaging over the Poisson traffic distribution leads to
\begin{equation}
\tpDL = \dlrate - \drop_1 \drop_2 \, \load(1-\peras)^2 e^{-\load(1-\peras^2)}.
\label{eq:tru_channelAgnostic}
\end{equation}

The channel-agnostic \ac{SFP} can then be optimised by setting $(\drop_1,\drop_2)$ to maximise $\tpDL$. To this aim we observe that, for any $\dlrate$, the maximum throughput $\tpDL^*$ is achieved when the loss factor expressed by the second addend in \eqref{eq:tru_channelAgnostic} is minimum. In other words, given an uplink configuration $(\load,\peras)$, we are interested in minimising the product $\drop_1 \drop_2$ under the constraint $\drop_1+\drop_2 \leq \dlrate/\tpSA$. The general solution is provided in App.~\ref{app:proofAgnostic}. From it, we infer the optimal settings $\drop_1^*=1$,  $\drop_2^*=\dlrate/\tpSA-1$, leading to
\begin{equation}
\tpDL^* = \dlrate \left ( 1-(1-\peras)\,e^{-\load \peras (1-\peras)} \right ) + \load(1-\peras)^2 e^{-\load(1-\peras^2)}
\label{eq:maxTru_channelAgnostic}
\end{equation}

\begin{figure}
\centering
\includegraphics[width=.6\columnwidth]{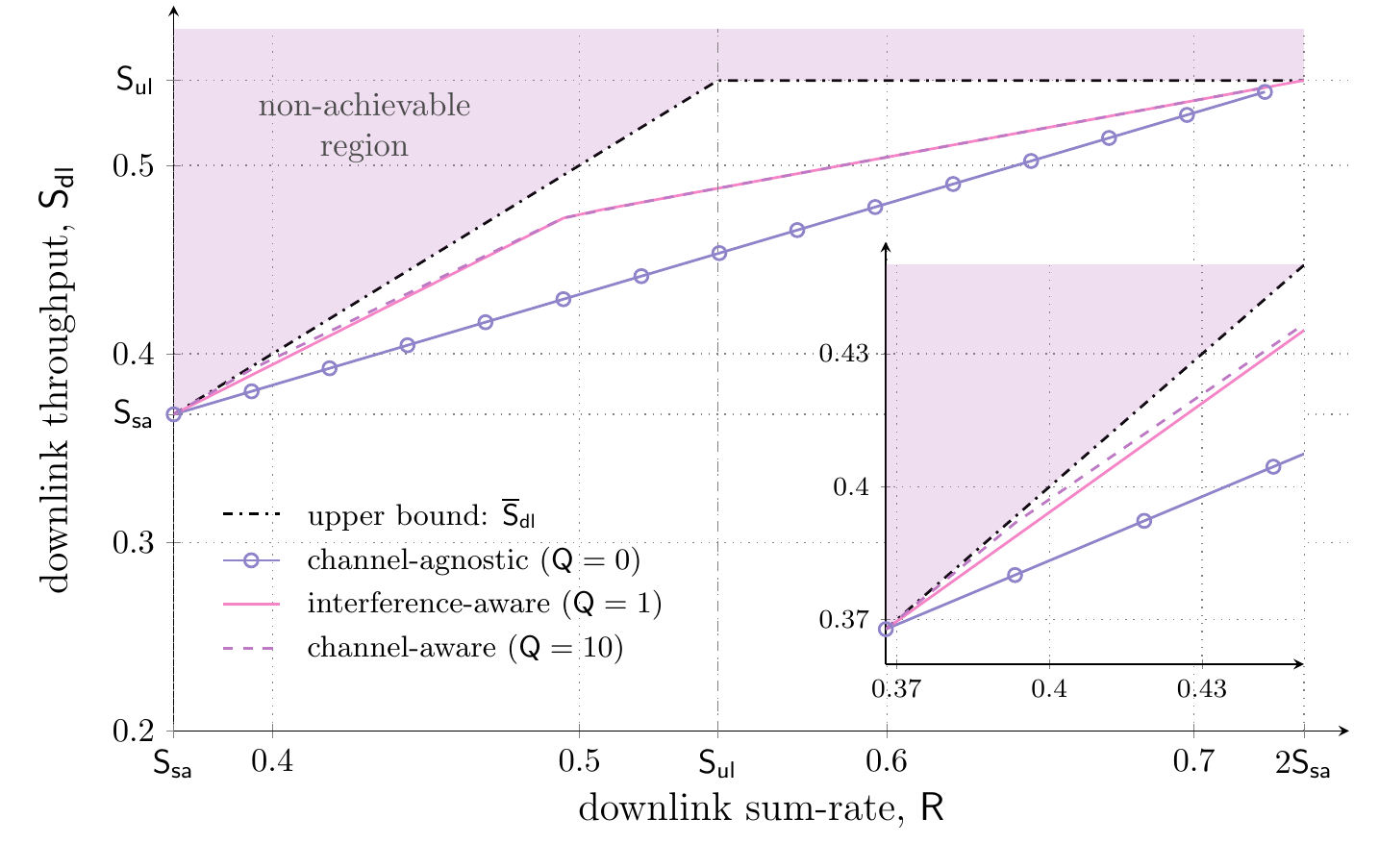}
\vspace{-.5em}
\caption{Downlink throughput $\tpDL$ vs. downlink sum-rate \dlrate\ for different simplified forwarding policies. $\peras=0.3$, $\load=1/(1-\peras).$} \,.
\label{fig:capacityRegionTheoretical}
\end{figure}

The obtained performance is summarised in Fig.~\ref{fig:capacityRegionTheoretical}, where the circle-marker line captures $\tpDL^*$ given the available sum-rate \dlrate\ in the exemplary case $\peras = 0.3$, $\load = 1/(1-\peras)$. The plot also reports (black dash-dotted line) the maximum achievable downlink throughput $\capDL(\dlrate)$ in \eqref{eq:Cdl}, which divides the plane in two regions and pinpoints values that can be aimed for. 
As expected, when each relay gets at least $\tpSA$ transmission opportunities per uplink slot, i.e. \mbox{$\dlrate \geq 2 \tpSA$}, $\tpDL = \tpUL$. On the other hand, the low-complexity of the \ac{SFP} is traded off for a larger downlink dimensioning. Indeed, the additional \mbox{$2\tpSA-\tpDL=\load(1-\peras)^2 e^{-\load(1-\peras^2)}$} resources needed to deliver all collected traffic to the sink in comparison to the \ac{RLC}-based approach stem from the possibility of having both relays transmit the same packet. Recalling the discussion of Sec.~\ref{sec:uplink}, the cost increases for smaller erasure rates, due to the more correlated decoding patterns experienced by the receivers.

It is also interesting to point out that, for lower values of \dlrate, the optimal allocation $(\drop_1^*,\drop_2^*)$ foresees one of the two relays to always forward all its incoming packets ($\dlrate_1=\tpSA$), triggering only partially the contribution of its fellow to limit inefficient transmissions of duplicate data units. From this standpoint, the policy represents a smart way to seamlessly take advantage of diversity in already deployed systems, as a single-relay scenario can be upgraded by plugging in an additional node and by incrementally allocating to it downlink bandwidth, without any change to the forwarding policy of the original receiver.

\subsection{Uplink channel aware simplified forwarding policies}

Let us now focus on policies that foresee relays make an educated choice on whether to drop or enqueue a data unit, based on the observation of what happened on the uplink channel. The intuition suggests that it is more likely for a packet to be retrieved by both receivers if it was the only one sent over the slot of interest, whereas the presence of several information units in an uplink slot reduces the chance for one of them to be decoded twice.  A reasonable approach is then to tune the buffering probability accordingly, to prevent duplicate transmissions in the downlink.

\subsubsection{Interference-aware simplified forwarding policies}
Consider first the simple case in which each collector bases its decision on whether the retrieved packet was the only one on the channel (i.e. absence of interference) or whether more than one user accessed the uplink channel over that slot  (i.e. presence of interference).  Note that, while this policy may appear elusive under the abstraction of on-off fading (no incoming power detected when a packet is erased), it turns out to be of high relevance in practical systems. Indeed, a data unit can often be decoded even in the event of a collision (e.g. by leveraging the capture effect \cite{Munari15:Massap}), and detecting the presence or absence of interference can rather easily be accomplished checking the noise level that affects the reception of the retrieved data unit.

Following the definition of \ac{SFP}, the case corresponds to having $\ms Q=1$, $\mc E_1 = \{1\}$ and $\mc E_2 = \mathbb N \backslash \{1\}$. In turn, the sum-rate can be expressed as
\begin{align}
\dlrate =& \left(\dropij{1}{1} + \dropij{2}{1}\right) \load (1-\peras)\, e^{-\load} \\
&+ \left(\dropij{1}{2} + \dropij{2}{2}\right)\,\sum\limits_{\users=2}^{\infty} \frac{\load^\users e^{-\load} }{\users!} \,\users\, (1-\peras) \peras^{\users-1}
\end{align}
where the first addend accounts for the case in which no interference was detected, whereas the summation considers all the situations in which at least one interfering packet affected the reception. As done for the channel-agnostic policy, we can compute the probability that a data unit is forwarded twice towards the sink, to obtain the downlink throughput
\begin{align}
\begin{split}
\tpDL =& \,\,\dlrate - \dropij{1}{1} \dropij{2}{1} \load (1-\peras)^2  e^{-\load} \\
&- \dropij{1}{2} \,\dropij{2}{2} \, \sum\limits_{\users=2}^{\infty} \frac{\load^\users e^{-\load} }{\users!} \,\users\, (1-\peras)^2 \peras^{2(\users-1)}.
\end{split}
\label{eq:truDL_interfAware_step0}
\end{align}

After simple  manipulations on \eqref{eq:truDL_interfAware_step0}, we can write the tackled optimisation problem as
\begin{align}
\begin{aligned}
\!\!\!\textrm{maximise } \quad &\tpDL = \dlrate - \coeffC\cdot \dropij{1}{1} \dropij{2}{1} - \coeffD \cdot \dropij{1}{2} \dropij{2}{2} \\
\!\!\!\textrm{s.t. }  \quad &\dlrate = \coeffA \cdot \left( \dropij{1}{1}  + \dropij{2}{1} \right) + \coeffB \cdot \left(\dropij{1}{2} + \dropij{2}{2} \right)
\end{aligned}
\label{eq:opt_problem_interfAware}
\end{align}
where ${\coeffC := \load (1-\peras )^2 e^{-\load}}$, $\coeffD := \load (1-\peras )^2 e^{-\load} ( e^{\load \peras^2} - 1 )$, $\coeffA := \load (1-\peras ) e^{-\load}$, and ${\coeffB := \load (1-\peras ) e^{-\load} \left( e^{\load \peras} - 1 \right)}$.
The solution is offered by the following result, whose proof is reported in App.~\ref{app:proofTheorem}:
\prop \label{prop:dropping} Let $\mathcal R_{1} = \left\{ \dlrate \,|\, \dlrate \in [ \tpSA, \coeffA+2\coeffB) \right\}$ and $\mathcal R_{2} = \left\{ \dlrate \,|\, \dlrate \in [ \coeffA+2\coeffB, 2 \tpSA] \right\}$. The maximum downlink throughput for the proposed interference-aware \ac{SFP} is
\begin{equation}
\tpDL^* =
\begin{cases}
\dlrate \, (1 - \coeffD/\coeffB) + \coeffD \,\tpSA/\coeffB &\text{for } \dlrate \in \mathcal R_1\\
\dlrate \, (1 - \coeffC/\coeffA)  + \coeffC (\tpSA + \coeffB )/\coeffA - \coeffD  &\text{for } \dlrate \in \mathcal R_2
\end{cases}
\label{eq:opt_interfAware}
\end{equation}
For $\dlrate \in \mathcal R_1$, the solution is obtained by setting $\dropVec_1^*= [ 1, 1 ]$ and  $\dropVec_2^* = [0, (\dlrate -\tpSA)/\coeffB]$ .  Conversely, for $\dlrate \in \mathcal R_2$, the  optimal working point is achieved for $\dropVec_1^* = [1, 1]$, and \mbox{$\dropVec_2^* = [(\dlrate-\tpSA-\coeffB)/\coeffA, =1]$}.

The resulting performance is once again reported in Fig.~\ref{fig:capacityRegionTheoretical} (solid line). In accordance to \eqref{eq:opt_interfAware}, the plot highlights two regions, both exhibiting a linear dependence of $\tpDL^*$ on $\dlrate$. In particular, when the overall downlink rate is lower than $\coeffA+2\coeffB$, the optimal allocation consists in having relay $1$ forward all the incoming traffic, whereas its fellow only delivers packets that were decoded in the uplink in the presence of interference. This confirms the intuition that data units collected when the \ac{SA} channel was accessed by more than one user bring a higher reward in terms of downlink throughput. On the other hand, when enough resources in the downlink are available for relay $2$ to deliver all such packets, the policy naturally enables it to gradually enqueue and transmit also information units collected in the absence of interference.  The higher probability for them to be duplicates of what forwarded by receiver $1$ is reflected in the lower slope of the throughput  curve in the rightmost region.

The figure clearly stresses the remarkable improvement unleashed by taking into account even partial information on the state of the uplink channel, proving how simple strategies can indeed provide performance that are not too far from the bound represented by \ac{RLC}. From this standpoint, two remarks are in order. In the first place, we notice that the switching point between $\mathcal R_1$ and $\mathcal R_2$ can be expressed as \mbox{$\coeffA+2\coeffB=2\tpSA - \load(1-\peras)e^{-\load}$}. Thus, for a given $\load$, higher erasure rates result in an extension of the region with higher throughput slope, further reducing the gap of the interference-aware \ac{SFP} with respect to $\capDL$. This trend highlights once more how stronger uncorrelation over the sets of decoded packets at the two relays may further benefit the class of proposed downlink strategies.
Secondly, while interference detection represents a practically viable basis to tune the buffering probabilities, the question on how more detailed side-information would impact the performance naturally arises.

\subsubsection{\ac{SFP} with complete channel awareness}
To tackle this, consider the ideal case in which both relays seamlessly and perfectly know how many packets were concurrently transmitted over an uplink slot. Under this hypothesis, we study a \ac{SFP} with $\mc E_j = \{j\}$, $j=1,\dots,\ms Q$ and \mbox{$\mc E_{\ms Q+1} = \mathbb N \backslash \{1,\dots,\ms Q\}$}.
The combinatorial approach followed so far can be employed to evaluate the buffer arrival rates $\dlrate_i$ and the  downlink throughput. After simple calculations, we get
\begin{equation}
\begin{split}
\dlrate  &=  \sum\limits_{j=1}^{\ms Q+1} \coeffA_j \left( \dropij{1}{j} + \dropij{2}{j} \right)\\[-.3em]
\tpDL &= \dlrate-  \sum\limits_{j=1}^{\ms Q+1}  \coeffB_j \, \dropij{1}{j}  \dropij{2}{j}
\end{split}
\label{eq:tru_channelAware}
\end{equation}
where the coefficients $\coeffA_j$ and $\coeffB_j$, ${j=1,\dots, \ms Q+1}$ are in \eqref{eq:coeffs_channelAware} at the bottom of the page and $\gamma(s,x)=\int_0^{x}t^{s-1}\, e^{-t} \, dt$ is the lower incomplete gamma function.

\begin{figure*}[!bp]
\normalsize
\hrulefill
\begin{equation}
\begin{aligned}
\begin{split}
\coeffA_i &= \load(1-\peras) e^{-\load} \cdot \frac{(\load \peras)^{i-1}}{(i-1)!}, \quad 1\leq i \leq \ms Q
&\hspace*{3em}  \coeffA_{\ms Q+1} &= \tpSA \, \gamma(\ms Q,\load \peras) / (\ms Q-1)!  \\
\coeffB_i &= \load(1-\peras)^2 e^{-\load} \cdot  \frac{(\load \peras^2)^{i-1}}{(i-1)!} , \quad 1\leq i \leq \ms Q
&\hspace*{3em}  \coeffB_{\ms Q+1} &= \load (1-\peras)^2 e^{-\load(1-\peras^2)} \, \gamma(\ms Q,\load \peras^2) / (\ms Q-1)! \\
\end{split}
\label{eq:coeffs_channelAware}
\end{aligned}
\end{equation}
\end{figure*}

Starting from \eqref{eq:tru_channelAware}, an optimisation problem analogous to the one in \eqref{eq:opt_problem_interfAware} can be stated, aiming at the buffering probabilities that maximise the downlink throughput for a given rate $\dlrate$. For arbitrary and potentially large values of $\ms Q$, however, an analytical solution is elusive. We thus follow a different approach, and conjecture that the idea underpinning the optimal working point for $\ms Q=1$ derived in Prop.~\ref{prop:dropping} extends to any value of $\ms Q$. More specifically, for the rate-region of interest, we let relay $1$ enqueue and forward all the received packets, setting $\dropij{1}{j} =1$, $\forall j$. On the other hand, when $\dlrate > \tpSA$, the second relay starts by buffering only the data units it receives that are less likely to have been decoded  by its fellow as well. This maps to $\dropij{2}{j}=0$ for $j\leq \ms Q$, while $\dropij{2}{\ms Q+1}$ grows linearly with the downlink rate. Eventually, $\dropij{2}{\ms Q+1}$ saturates to $1$ when $\dlrate = \tpSA + \coeffA_{\ms Q+1} = \tpSA \left[ 1 + \gamma(\ms Q,\load \peras) / (\ms Q-1)! \right]$.
After this point, additional resources allocated to relay $2$ will be used to store and forward packets received over slots accessed by $\ms Q$ users in the uplink.  Following the same reasoning, $\dropij{2}{\ms Q}$ is then linearly increased until it reaches one, i.e. for $\dlrate = \tpSA + \coeffA_{\ms Q+1} + \coeffA_{\ms Q}$. Iterating this approach, $\ms Q+1$ rate regions can be identified, where the second receiver is progressively allowed to deliver information units which are more likely to be duplicates of what forwarded by its fellow. More formally, we state the following:
\conj \label{conj1}
For a channel aware \ac{SFP}, let the rate values $r_{\ms Q+1}  < \dots < r_1$ be defined as  $r_{\ms Q+1} = \tpSA \left[ 1 + \gamma(\ms Q,\load \peras)/(\ms Q-1)! \right]$ and
\begin{equation}
 r_i = \tpSA \left( 1  + \frac{\gamma(i,\load \peras) - e^{-\load \peras} (\load \peras)^{i-1}}{(i-1)!} \right), \, 1\leq i \leq \ms Q \,.
\end{equation}
Accordingly, introduce the $\ms Q+1$ disjoint downlink rate regions $\mc R_{\ms Q+1} = \left\{ \dlrate \in [ \tpSA, r_{\ms Q+1}) \right\}$,  $\mc R_i = \left\{ \dlrate \in [ r_{i+1}, r_{i}) \right\}$, ${1\leq i \leq \ms Q}$.
We conjecture that, for any $\dlrate \in \mc R_n$, $n\in \{1,\dots,\ms Q+1\}$, the buffering probabilities maximising the downlink throughput are $\dropij{1}{j}^* = 1$, $\forall j$, and
\begin{align}
\dropij{2}{j}^*&=
\begin{cases}
\,\,1 \,\, & n < j \leq \ms Q+1 \\
\, \left(\dlrate -\tpSA - \sum_{i=n+1}^{\ms Q+1}\coeffA_i \right) / \coeffA_n \,\, & j=n\\
\,\, 0  \,\, &j < n \\
\end{cases}
\label{eq:optQ_channelAware}
\end{align}
Under this choice, letting \mbox{$\coeffE_i := \coeffB_i/\coeffA_i$}, we have
\begin{equation}
\tpDL^* \!=\!
\begin{cases}
\left(1-\coeffE_{\ms Q+1} \right) \dlrate + \tpSA \coeffE_{\ms Q+1}  &  \!\dlrate \in \mathcal R_{Q+1} \\[2mm]
\left( 1 - \coeffE_i \right) \dlrate + \left[ 1 + \gamma(i,\load \peras) / (i-1)! \right] \tpSA \coeffE_i &  \! \dlrate \in \mathcal R_{i\leq Q}\\[1mm]
\,\, - \load (1-\peras)^2 e^{-\load(1-\peras^2)}\, \gamma(i,\load \peras^2) / (i-1)!
  &
\end{cases}
\label{eq:opt_channelAware}
\end{equation}

The derivation of \eqref{eq:opt_channelAware}, although not reported here due to space constraints, follows directly by some manipulation after plugging the probability values of \eqref{eq:optQ_channelAware} into \eqref{eq:tru_channelAware}. The accuracy of the conjecture has been verified by means of constrained numerical optimisation techniques applied to \eqref{eq:tru_channelAware} for a variety of uplink configurations $(\load,\peras)$, always obtaining values in excellent agreement with the presented analytical expressions.
Leaning on this result, we report in Fig.~\ref{fig:capacityRegionTheoretical} the achievable throughput against the downlink rate when a very accurate knowledge of the uplink channel conditions in terms of size of the collision sets is available at relays, i.e. $\ms Q = 10$ (dashed line). An inspection of the $r_i$ introduced in the conjecture reveals how the starting point of the rightmost region $\mathcal R_1$ (and the downlink throughput achieved therein) does not vary with $\ms Q$. Increasing the level of channel knowledge, thus, leads to a larger number of smaller partitions of $\tpSA \leq \dlrate \leq \dlrate_1^*$. On the other hand, such leftmost regions are precisely the ones characterised by a stronger slope of the throughput curve, earned leveraging additional side information. The combination of the two effects significantly curbs the benefits brought by a more accurate knowledge of the number of users accessing the uplink channel. This is clearly highlighted in the plot, where the $\ms Q=10$ curve exhibits a trend which is very close to the one of its $\ms Q=1$ counterpart, with a limited gain only in the downlink rate region which is in fact of less interest for multi-receiver systems (i.e., when the total available rate is slightly larger than the one necessary to collect the traffic of a single relay). Such a result is remarkable, and suggests how a simple and practically viable strategy which makes forwarding decisions only based on interference detection can indeed reap a noticeable fraction of the downlink throughput  achievable by means of a large family \ac{SFP}, offering performance not too far away from the ones of \ac{RLC}.

\section{Impact of Finite Buffer Size on\\Downlink Strategies}
\label{sec:buffer_downlink}

The downlink study carried out in the previous sections relies on having infinite-size queues available at the relays. Besides being paramount for analytical tractability, this common modelling assumption is key to many reported results. Indeed, the \ac{RLC} ability to deliver all collected information using the minimum downlink rate requires relays to buffer received packets for an asymptotically long interval, prior to efficiently encoding them. Similarly, the eventual transmission towards the sink of all enqueued data units that characterised \ac{SFP}s only holds if no overflow condition can arise.

However, in many practical settings, constraints in terms of hardware memory or application-driven latency may dictate the use of short buffers, departing from the modelling approach tackled so far. To further understand the effectiveness of the proposed downlink policies, we thus complement our investigation relaxing the assumption of unbounded relay queues, with focus on the $\nrx = 2$ case.

\subsection{Random linear coding with finite-size buffers}

Let us consider initially the \ac{RLC} approach, and assume that receivers equipped with a buffer of size \numulslot\ listen to the uplink channel for \numulslot\ slots.  At the end of the observation period, the two relays generate respectively $\collK{1} := \numulslot \dlrate_1$ and $\collK{2} := \numulslot \dlrate_2$ linear combinations over $\mathbb F_\fieldOrd$ of what collected, and forward them over the downlink. Following the methodology and notation presented in Sec.~\ref{sec:nc}, the sink can then obtain the reduced system of linear equations reported in \eqref{eq:exampleG_K=2}.
For the sake of clarity, we recall that $\MxCombRedK{1,\{1\}}$ is the \mbox{$(\collK{1} \times \linllK{1})$} matrix with elements in $\mathbb{F}_{\fieldOrd}$ collecting the coefficients applied to linear combinations involving the $\linllK{1}$ packets decoded by relay $1$ only. Likewise, the $(\collK{2}\times \linllK{2})$ matrix $\MxCombRedK{2,\{2\}}$ weighs the $\linllK{2}$ data units received solely by relay $2$. Finally, the $(\collK{1}\times \linllK{1,2})$ $\MxCombRedK{1,\{1,2\}}$ and $(\collK{2} \times \linllK{1,2})$ $\MxCombRedK{2,\{1,2\}}$ matrices group coefficients of the first and second receiver involving the $\linllK{1,2}$ packets decoded by both.

In order to compute the number of information units retrieved at the sink, Gauss-Jordan elimination is applied. After some line and column reordering, the procedure leads to a useful reformulation of the coefficient matrix as\\[-2mm]
\begin{align}
\MxCombRed' = \kbordermatrix{
                & \rankK{1}     & \rankK{2}         & \rankK{1,2}           &            & \mxColK{1}    & \mxColK{2}    & \mxColK{1,2} \\
\rankK{1}       &\Id_{\rankK{1}}& \bm{0}            & \bm{0}                & \vrule     & \MxA          & \bm{0}        & \MxUone \\
\rankK{2}       & \bm{0}        & \Id_{\rankK{2}}   & \bm{0}                &\vrule& \bm{0}        & \MxB          & \MxUtwo \\
\rankK{1,2}     & \bm{0}        & \bm{0}            & \Id_{\rankK{1,2}}     &\vrule& \bm{0}        & \bm{0}        & \MxL \\[.1em]
\cline{2-8}
                & \bm{0}        & \bm{0}            & \bm{0}                &\vrule& \bm{0}        & \bm{0}        & \bm{0}
}\,.
\label{eq:Mx_second_two}
\end{align}\\
Here, $\Id_x$ refers to the identity matrix of size $x$, while $\rankK{1}$, $\rankK{2}$ represent the rank of the matrices $\MxCombRedK{1,\{1\}}$ and $\MxCombRedK{2,\{2\}}$. For ease of compactness, let us furthermore denote $\mxColK{1}:=\linllK{1}-\rankK{1}$, $\mxColK{2}:=\linllK{2}-\rankK{2}$ and $\mxColK{1,2}:=\linllK{1,2}-\rankK{1,2}$. Following this notation, $\MxA$ and $\MxB$ have size $(\rankK{1} \times \mxColK{1})$ and $(\rankK{2}\times \mxColK{2})$ respectively, and collect all non-zero elements left after Gauss-Jordan elimination on the matrices $\MxCombRedK{1,\{1\}}$ and $\MxCombRedK{2,\{2\}}$.  In turn, sub-matrices $\MxUone$ and $\MxUtwo$ are of dimension $(\rankK{2}\times \mxColK{1,2})$ and $(\rankK{2} \times \mxColK{1,2})$. Finally, $\MxL$ is a $(\rankK{1,2},\times \mxColK{1,2})$ matrix, where $\rankK{1,2}\leq \left[(\collK{1} + \collK{2}) - (\rankK{1} + \rankK{2})\right]$.
Clearly, all the introduced sizes are stochastic quantities, as they depend on the reception patterns experienced at the relays over the \numulslot\ uplink slots as well as on the generated linear combinations.

Let now $\collSinkone \leq \linllK{1}$, $\collSinktwo \leq \linllK{2}$ and $\collSinkboth\leq \linllK{1,2}$ be the number of data units obtained by the sink that were originally decoded in the uplink only by relay $1$, only by relay $2$ and by both of them, respectively. Indicating as $\collSinkoneRV$, $\collSinktwoRV$ and $\collSinkbothRV$ the corresponding random variables, the downlink throughput can be expressed as
\begin{equation}
\label{eq:th_dw}
\tpDL = \frac{1}{\numulslot} \left(\,\mathbb{E}[\collSinkoneRV] + \mathbb{E}[\collSinktwoRV] + \mathbb{E}[\collSinkbothRV] \,\right).
\end{equation}

Let us focus initially on data units that were only available at relay $1$, restricting our attention to the first $\rankK{1}$ rows of the coefficient matrix. After the performed Gauss-Jordan elimination, the sink can retrieve one such packet only if a single non-zero component appears in a row of $\MxCombRed'$, i.e. if all the elements in the corresponding rows of $\MxA$ and $\MxUone$ are null. For a specific realisation of the coefficient matrix, the probability of such event can be approximated as $\fieldOrd^{-(\mxColK{1}+\mxColK{1,2})}$, leading to a binomial distribution for $\collSinkoneRV$  with parameters $(\rankK{1},\fieldOrd^{-(\mxColK{1}+\mxColK{1,2})})$.
A similar reasoning can be applied to packets only available at relay $2$ and at both receivers, obtaining that, conditioned on the realisation of $\MxCombRed'$,  $\collSinktwoRV \sim {\rm{Bin}}(\rankK{2}, \fieldOrd^{-(\mxColK{2} + \mxColK{1,2})})$ and $\collSinkbothRV \sim {\rm{Bin}}(\rankK{1,2}, \fieldOrd^{-\mxColK{1,2}})$.

Leaning on this, by the law of total expectation \eqref{eq:th_dw} can be approximated as
\begin{equation}
\begin{aligned}
\label{eq:exp_c1}
\tpDL \simeq \frac{1}{\numulslot} & \sum_{\substack{ \linllK{1}, \linllK{2}, \linllK{1,2} \\
\rankK{1}, \rankK{2}, \rankK{1,2} }} \left( \frac{\rankK{1}}{\fieldOrd^{(\mxColK{1}+\mxColK{1,2})}} +
\frac{\rankK{2}}{\fieldOrd^{(\mxColK{2}+\mxColK{1,2})}} + \frac{\rankK{1,2}}{\fieldOrd^{\mxColK{1,2}}} \right) \\
&\cdot P( \rankK{1} \,|\, \linllK{1} )\,  P( \rankK{2} \,|\, \linllK{2} )\, P( \rankK{1,2} \,|\, \linllK{1,2}, \rankK{1}, \rankK{2} )  \\
& \cdot P( \linllK{1}, \linllK{2}, \linllK{1,2} )
\end{aligned}
\end{equation}
where we indicate for generic r.v. $X$ and $Y$ \mbox{$\pr\{X=x | Y = y\}$} as $P(x|y)$, and $\pr\{X=x, Y = y\}$ as $P(x,y)$.
To compute the downlink throughput it is thus necessary to derive the conditional probabilities of the sub-matrix ranks as well as the joint probability mass function of the uplink packets received by the two relays.

We tackle initially the former task, with focus on $P(\rankK{1} | \linllK{1})$. The probability that a generic matrix with elements drawn uniformly at random in $\mathbb{F}_{\fieldOrd}$ has rank $\rankK{1}$, given that is constituted by $\collK{1}$ rows and $\linllK{1}$ columns, can be recursively computed \cite{Lidl96_FiniteFields}.
To this aim, we first consider the prototype $(1 \times \linllK{1})$ matrix $\MxM \in \mathbb{F}_{\fieldOrd}^{\{1 \times \linllK{1}\}}$. The probability that $\MxM$ has rank $0$ is $\fieldOrd^{-\linllK{1}}$, whereas it has rank $1$ with probability $(1-\fieldOrd^{-\linllK{1}})$. Now, for a generic $\collK{1}$, the probability that $\MxM \in \mathbb{F}_{\fieldOrd}^{\{\collK{1} \times \linllK{1}\}}$ has rank $\rankK{1}$ is
\begin{equation}
\label{eq:mx_recursion}
\begin{aligned}
&P(\rankK{1}|\linllK{1}) =\fieldOrd^{(\rankK{1}-\linllK{1})} \,\pr \left\{\rank\left( \MxM \in \mathbb{F}_{\fieldOrd}^{\{\collK{1}-1 \times \linllK{1}\}} \right)=\rankK{1}\right\}+\\
&\left(1- \fieldOrd^{(\rankK{1}-\linllK{1}-1)}\right) \pr\left\{\rank\left( \MxM\in \mathbb{F}_{\fieldOrd}^{\{\collK{1}-1 \times \linllK{1}\}} \right)=\rankK{1}-1\right\}\\
& \qquad \qquad \text{with $\rankK{1}=0,1,...,\min(\collK{1},\linllK{1})$.}
\end{aligned}
\end{equation}\\[-1mm]
The result straightforwardly applies to $P(\rankK{2}| \linllK{2})$. Similarly, $P(\rankK{1,2}|\linllK{1},\rankK{1},\rankK{2})$ can be computed referring to a matrix with $\collK{1}+\collK{2}-(\rankK{1}+\rankK{2})$ rows (i.e. the number of equations that do not involve any packet available only at one relay), and $\linllK{1,2}$ columns (i.e. the number of packets that are indeed received by both).

On the other hand, the joint probability mass function $P(\linllK{1},\linllK{2}, \linllK{1,2})$ can be tracked effectively via a homogeneous Markov chain. More precisely, let \mbox{$ [ \linllK{1}(\slot), \linllK{2}(\slot), \linllK{1,2}(\slot) ]$} be the chain state at the start of the $\slot$-th observed uplink slot, tracking the number of packets received so far solely by the first, solely by the second, and by both relays, respectively. Moreover, denote for compactness  as $\st_{\linComb,\linComb'}$ the probability of transitioning from state $\linComb$ at slot $m$ to state $\linComb'$ at slot $m+1$, and the three standard basis vectors of $\mathbb N^3$ as $\eo=[1,0,0]$, $\et=[0,1,0]$ and $\etr=[0,0,1]$.  Following this notation, each time unit can see five possible transitions for the chain, whose probabilities follow by simple combinatorial arguments similar to the ones discussed in Sec.~\ref{sec:uplink}:
\begin{equation}
\begin{aligned}
\st_{\linComb,\linComb} &= 1 - \tpUL + \Tx\\
\st_{\linComb,\linComb +\eo} = \st_{\linComb,\linComb+\et} &= \tpUL - ( \tpSA + \Tx ) \\
\st_{\linComb, \linComb+\etr} &= 2\,\tpSA - \tpUL \\
\st_{\linComb, \linComb+\eo+\et} &=  \Tx \\
\end{aligned}
\label{eq:markov_transitions}
\end{equation}\\[-2mm]
where $\tpUL$ follows from \eqref{eq:truUplink} setting $\nrx=2$, $\tpSA$ is given by \eqref{eq:Tsa}, and the ancillary quantity $\psi={ (\load \peras)^2 (1-\peras)^2 \, e^{-\load (1-\peras^2)}}$ captures the probability that the same packet is decoded at both relays.
The values in \eqref{eq:markov_transitions} uniquely identify the transition matrix for the Markov chain, so that $P(\linllK{1},\linllK{2}, \linllK{1,2})$ follows as its $\numulslot$-th step evolution when forcing the initial state as $[0,0,0]$.
Plugging the derived probabilities for the ranks and the collected packets into \eqref{eq:exp_c1} allows to have an analytical calculation of the average downlink throughput, for any rate allocation and observation interval \numulslot.

\begin{figure}
\centering
\includegraphics[width=.6\columnwidth]{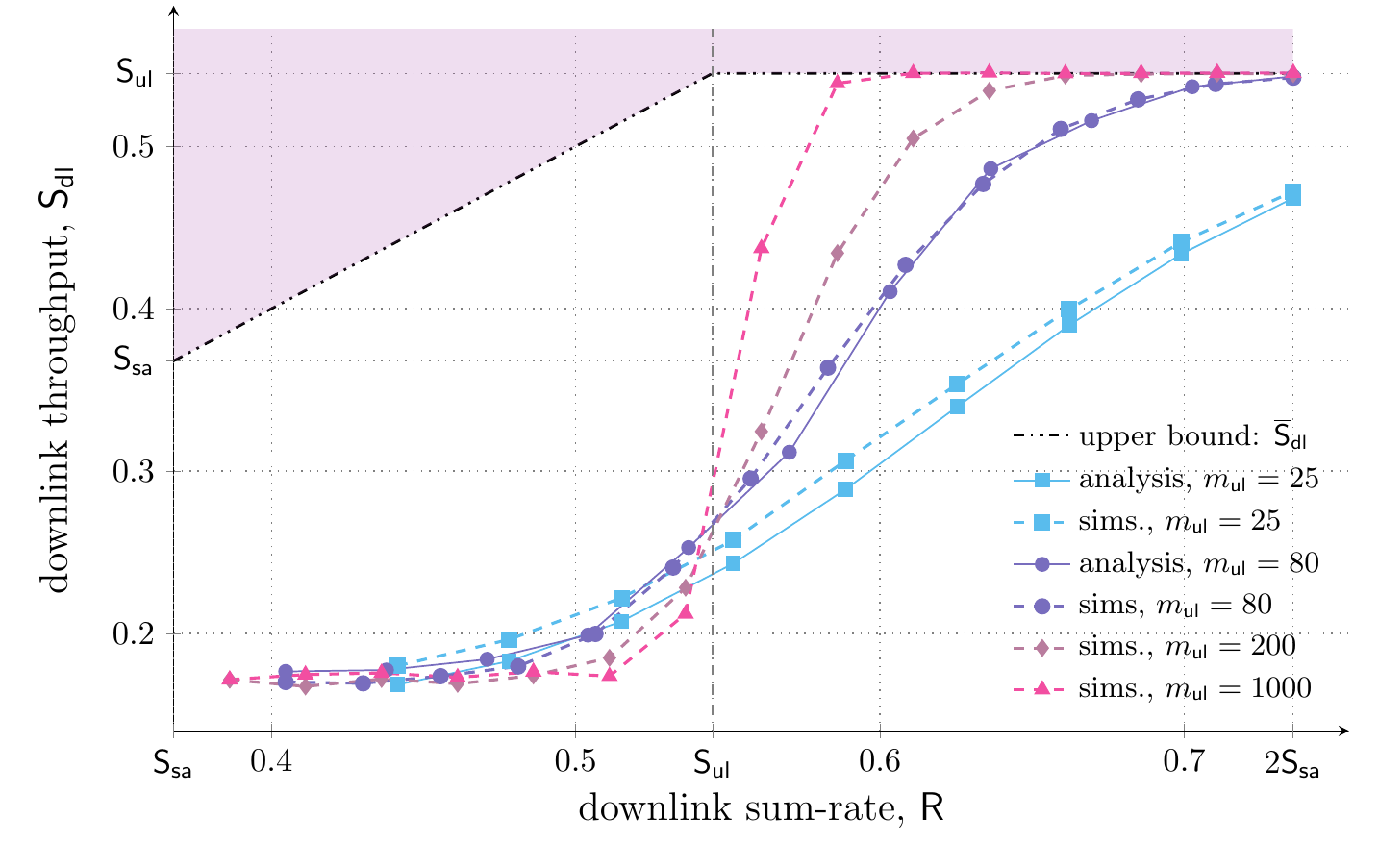}
\vspace{-.5em}
\caption{Downlink throughput vs sum-rate achieved by a \ac{RLC} forwarding policy under finite buffer size. $\nrx =2$ relays are considered, and the uplink is operated with $\peras=0.3$ and $\load =1/(1-\peras)$.}
\label{fig:NC_finite}
\end{figure}

To gauge the impact of finite buffer size, let us then consider the case in which the downlink is na\"{i}vely tuned assuming maximum efficiency for \ac{RLC} (i.e. $\numulslot \rightarrow \infty$). Instantiating the results of Prop.~\ref{prop:rlcAchiev} for the $\nrx=2$ case, when $\dlrate \geq \tpUL$ this translates into granting one relay a rate equal to $\tpSA$ and to its fellow the remaining resources.  We generalise this approach by setting \mbox{$\dlrate_1 = \tpSA$} and \mbox{$\dlrate_2 = \dlrate - \tpSA$} for any sum-rate. The analytical performance in terms of downlink throughput achieved using such rates for different values of \numulslot\ is reported against $\dlrate$ by solid lines in Fig.~\ref{fig:NC_finite}, assuming $\peras = 0.3$ and $\load = 1/(1-\peras)$. All reported trends assume linear combinations to be performed over $\mathbb F_2$.\footnote{Additional results, not reported due to space constraints, revealed a limited impact of the field order \fieldOrd.}
The accuracy of the developed framework was also verified by means of Monte Carlo simulations, implementing Poisson traffic over the on-off fading uplink channel, generation of linear combinations at the relays and Gauss-Jordan elimination at the sink to retrieve as many data units as possible. The outcome is reported by dashed lines in the plot, buttressing the analytical trends with a tight match. In view of this, to ease readability, only simulation results are depicted for larger values of \numulslot.

Fig.~\ref{fig:NC_finite} offers several relevant insights. In the first place, a drastic performance degradation compared to the asymptotic upper bound is experienced for $\numulslot = 25$. Remarkably, delivery of all collected packets to the sink ($\tpDL = \tpUL$) cannot be achieved even when the resources allotted to each of the relays matches the average uplink throughput it experiences (i.e. for $\dlrate = 2\tpSA$). Such a trend reflects the considered rate assignment, which is based on \emph{average} uplink performance. Indeed, transmissions opportunities granted to relay $2$ are tuned for it to efficiently encode   the statistically expected $(\tpUL-\tpSA) \numulslot$ packets that were not decoded by its fellow. When $\numulslot$ is limited, however, stochastic fluctuations may often bring to collecting a number of data units that largely differs from this value, leading to suboptimal performance.

As the buffer size increases, a threshold behaviour starts to emerge. For $\dlrate < \tpUL$,  small values of $\tpDL$ are achieved. 
In this region, even if full-rank sub-matrices are obtained, too few equations are available, hindering packet retrieval.
Therefore, the downlink throughput is bounded away from the upper bound offered by $\capDL$, regardless of \numulslot. Conversely, when the granted sum-rate exceeds the threshold $\tpUL$, the size of queues available at the relays starts to have a significant effect.  Already for $\numulslot=80$, all collected traffic can be delivered over the downlink when \dlrate\ approaches $2 \tpSA$. Allowing larger observation periods, the minimum downlink rate required to fulfil the task further decreases, as exemplified by the $\numulslot=1000$ case, where few additional resources with respect to the asymptotical limit $\tpUL$ are required to reach the maximum downlink throughput.

The reported results clearly highlight how care shall be taken when implementing a \ac{RLC} forwarding policy in practical system. On the one hand, a minimum amount of downlink resources shall be allotted for the efficiency of the encoding strategy to kick in. Secondly, when limited buffering capabilities or strong delay constraints are to be faced, rate allocations among the relays that depart from the optimal asymptotic solution shall be considered. Such a remark paves road for further relevant studies, which we regard as part of our future work.


\subsection{Simplified forwarding policies with finite-size buffers}

\begin{figure}
\centering
\includegraphics[width=.6\columnwidth]{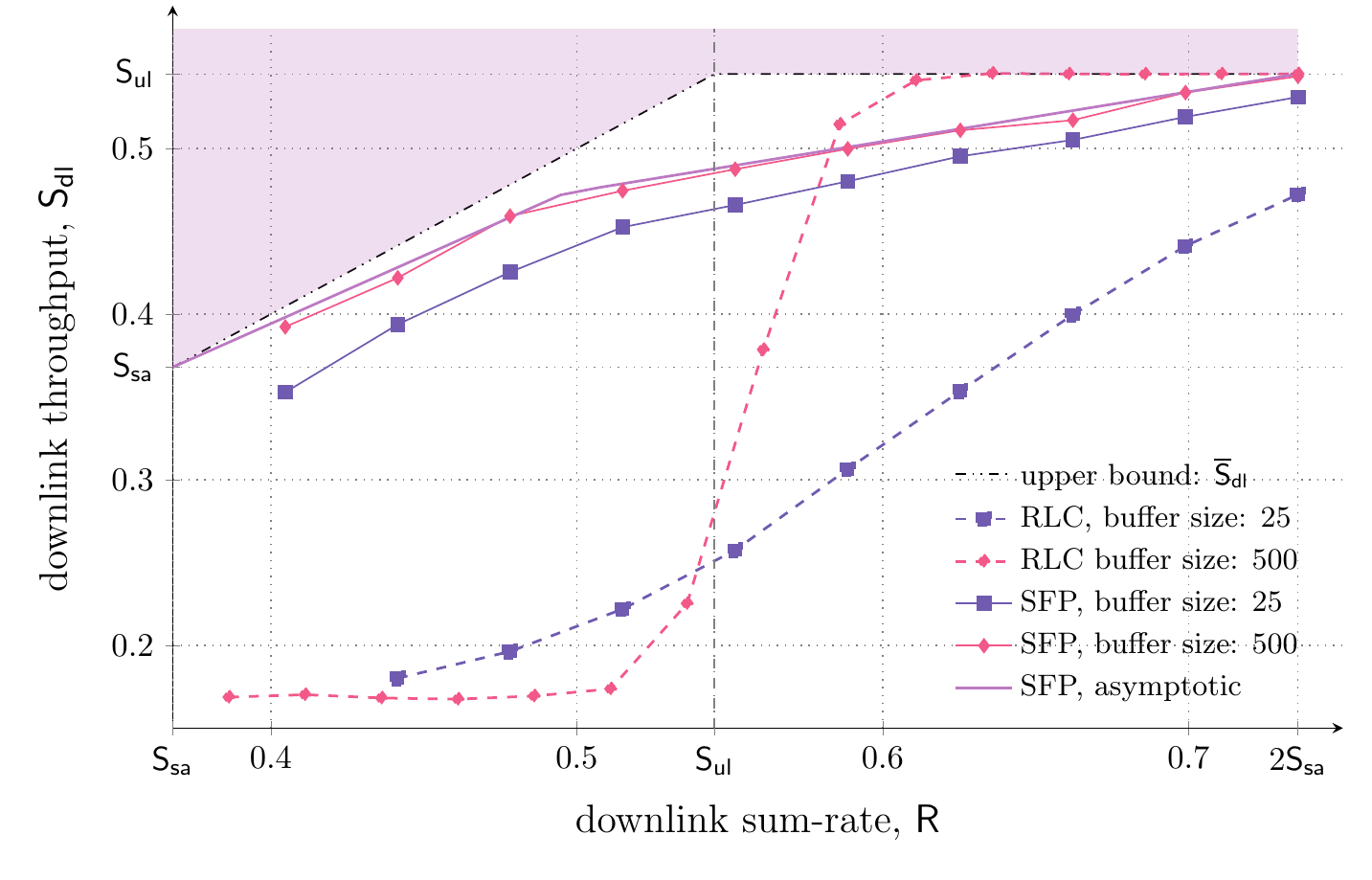}
\vspace{-.5em}
\caption{Downlink throughput vs sum-rate for the interference-aware \ac{SFP} and \ac{RLC} with finite-sized buffers. $\nrx =2$ relays are considered, and the uplink is operated with $\peras=0.3$ and $\load =1/(1-\peras)$.}
\label{fig:Comparison2}
\end{figure}

We conclude our discussion studying the impact finite buffers play on the simplified strategies introduced in Sec.~\ref{sec:downlinkSimplified}. From the analytical standpoint, the problem can be tackled resorting to discrete-time queueing theory tools, describing each relay as a Geo/D/$1$ system \cite{Gravey90}. While viable, however, the modelling approach soon becomes analytically cumbersome. We thus rely on Monte Carlo simulations, and focus on the \emph{interference-aware} \ac{SFP}, whose enqueueing probabilities are once again tuned for the $\numulslot \rightarrow \infty$ case. Two buffer sizes are considered, namely $25$ and $500$, to capture settings ranging from stringent to very relaxed delay constraints. Results are shown in Fig.~\ref{fig:Comparison2} together with the asymptotic performance that was derived analytically. For the sake of comparison, the behaviour of \ac{RLC} is reported as well.

When short queues are considered (i.e. buffer size $25$), the \ac{SFP} exhibits a loss with respect to the unbounded case, as decoded packets may be dropped not only due to an educated decision made by the relay but to overflow conditions as well. The performance decrease is however rather contained ($\sim 5 \%$), and quickly vanishes when larger queues are available. It is also relevant to observe how, for a buffer of $25$ data units, the simpler policy outperforms \ac{RLC} for the entire range of reported rates, with gains exceeding $100$\%. The trend is confirmed also for larger buffers and low sum-rates, stressing how the introduced \ac{SFP} may be a better choice when few resources are available in the downlink. On the other hand, the benefits of packet-level coding become apparent for $\tpUL <\dlrate \leq 2\tpSA$. In such conditions, \ac{RLC} can deliver the whole collected uplink throughput, whereas the simplified strategies inherently drop some of the collected packets regardless of the queue size.  

\section{Conclusions} \label{sec:conclusions}
This paper investigated the performance of a wireless network where users transmit following a slotted ALOHA policy towards a set of non-cooperative receivers, which, in turn, forward information towards a sink. Considering an on-off fading model for the user-receiver channels, we offered exact analytical expressions for the uplink performance for an arbitrary number of receivers. Although beneficial in terms of throughput and packet loss rate, an increase in the number of relays poses challenges in terms of downlink dimensioning. To provide guidance on this task, we derived a bound on the minimum amount of resources required to deliver all information collected at the relays with arbitrarily low error probability, and showed its achievability via random linear coding. Moreover, we introduced simplified and practical forwarding strategies that require no packet-level coding, and analysed their performance. Finally, the behaviour of random linear coding downlink policies was modelled analytically in the presence of finite buffers at relays (i.e. in the presence of latency constraints), and compared with the proposed simplified strategies. Results show that random linear coding is heavily sub-optimal for stringent buffer sizes, or when the downlink resources fall below the uplink throughput. Indeed, in such scenarios, simplified strategies can offer $2$-fold or higher throughput gains. For large buffer sizes (or less stringent delay constraints), packet-level coding becomes the preferred choice if the throughput has to be maximised.

\appendices
\section{} \label{app:achievability}

According to the notation in the proof of Prop.~\ref{prop:rlcAchiev}, we derive that
$\sum_{\LSet \subseteq \rxSubSet}  \left|\,\USet_{\LSet}\right| = \left|\bigcup_{\rx \in \rxSubSet} \setArx \backslash \bigcup_{\rx \in \overline{\rxSubSet}} \setArx \right|$. To this aim, we lean on the following:
\begin{lemma}
 For a collection of sets $\hSet_1,\hSet_2,\ldots, \hSet_\nrx$ and a subset $\rxSubSet\subseteq \{1,\ldots,\nrx\}$
\begin{align}
 \bigcup_{\rx \in \rxSubSet} \hSet_\rx = \bigcup_{\LSet \subseteq \rxSubSet}\left( \bigcap_{l\in \LSet} \hSet_l \backslash \bigcup_{j \in \rxSubSet\backslash \LSet} \hSet_j\right),
\end{align}
where the sets on the RHS do not intersect and thus form a partition of the LHS.
\end{lemma}
\begin{IEEEproof}
 We first show that any element $s\in  \bigcup_{\rx \in \rxSubSet} \hSet_\rx$ is also included in the RHS.  Assume $\LSet$ is the subset of largest cardinality such that $s \in \hSet_l,~\forall l \in \LSet$. Clearly, $s \in \bigcap_{l\in\LSet}\hSet_l$ but $s \not \in \bigcup_{j\in \rxSubSet\backslash \LSet} \hSet_j$. It follows that $s \in \bigcap_{l\in\LSet}\hSet_l \backslash \bigcup_{j\in \rxSubSet\backslash \LSet} \hSet_j$. This is true for some subset $\LSet \subseteq \rxSubSet$.
Second, we show that this subset is unique. Let again $\LSet$ be the subset of largest cardinality such that $s \in \hSet_l,~\forall l \in \LSet$ and choose a different subset $\vSet\subseteq \rxSubSet$, $\vSet \not = \LSet$. Then, either $s \not \in \bigcap_{l \in \LSet} \hSet_l$ or $s\in \bigcup_{j \in \rxSubSet \backslash \LSet} \hSet_j$. The element $s$ is thus only included in $\bigcap_{l\in\LSet}\hSet_l \backslash \bigcup_{j\in \rxSubSet\backslash \LSet} \hSet_j$.
\end{IEEEproof}
By choosing $\hSet_\rx = \setArx \backslash \bigcup_{\rx \in \overline{\rxSubSet}} \setArx$, the sought result follows by elementary set operations.
\section{}  \label{app:proofAgnostic}

\begin{lemma}
For any $\alpha \in [0,2]$, let $x$ and $y$ be real numbers such that $x,y \in [0,1]$ and $x+y= \alpha$.
Then, the $\{x,y\}$ pairs that minimise the product $z= xy$ are given by
\begin{align}
\textrm{for } \alpha\in[0,1]:& \quad \{x=0, y= \alpha \} \textrm{ or, } \{x=\alpha, y=0\} \\
\textrm{for } \alpha\in(1,2]:& \quad \{x=1, y= \alpha-1\} \textrm{ or, } \{x=\alpha-1, y=1\}
\end{align}
\label{lemma:dropping}
\end{lemma}

\vspace{-6mm}
\begin{IEEEproof}
Writing $z$ as a function of $x$, we get $z=-x^2+\alpha x$, which represents a parabola with downward concavity and zeroes for $x=0$ and $x=\alpha$. Furthermore, imposing the conditions $x\in[0,1]$ and $\alpha-x=y\in[0,1]$, the region of interest restricts to $\max\{0,\alpha-1\}\leq x\leq \min \{ \alpha, 1 \}$. When $\alpha\leq 1$, the minimum value of $z$ in the studied domain is thus reached either when $x=0$ or $x=\alpha$, while, for $\alpha\in(1,2]$ the minimum value is obtained for $x=\alpha-1$ and $x=1$. The values of $y$ follow immediately.
\end{IEEEproof}

\section{Proof of Proposition~\ref{prop:dropping}} \label{app:proofTheorem}
As a preliminary remark, note that all the coefficients $\coeffA,\coeffB,\coeffC,\coeffD$ in \eqref{eq:opt_problem_interfAware} are strictly positive, and that, for any admissible values of $\load$ and $\peras$, the inequality $\coeffC/\coeffA > \coeffD/\coeffB$ holds. For the sake of a simplified notation, let $x_1:=\dropij{1}{1}$, $x_2:= \dropij{2}{1}$, $y_1:=\dropij{1}{2}$, and $y_2:=\dropij{1}{2}$. Furthermore, let us introduce $\alpha = x_1+y_1$ and $\beta = x_2 + y_2$, with $\alpha,\, \beta \in [0,2]$, as well as the auxiliary function $f(\mathbf x) = \coeffC\, (x_1 y_1) + \coeffD\,(x_2 y_2)$, where $\mathbf x =[x_1,x_2,y_1,y_2]$ and  $f(\mathbf x) \geq 0$. We are then interested in maximising $\tp = \dlrate - f(\mathbf x)$ subject to $\dlrate = \alpha \coeffA + \beta \coeffB$ or, equivalently in minimising $f(\mathbf x)$ under the same constraint. Let us notice that the first addend of $f(\mathbf{x})$ only contains the variables that determine $\alpha$, while the second addend of $f(\mathbf{x})$ solely defines the value of $\beta$. It is then possible to solve the optimisation problem by considering four non-overlapping regions: $\mathcal G_1 = \{\mathbf x \, | \, \alpha \in [0,1], \beta \in [0,1] \}$, $\mathcal G_2 = \{\mathbf x \, | \, \alpha \in [1,2], \beta \in [0,1] \}$, $\mathcal G_3 = \{\mathbf x \, | \, \alpha \in [0,1], \beta \in [1,2] \}$, $\mathcal G_4 = \{\mathbf x \, | \, \alpha \in [1,2], \beta \in [1,2] \}$. $\mathcal G_1$ can immediately be discarded as $\alpha,\beta < 1$ implies $\dlrate < \coeffA+\coeffB = \tpSA$, identifying a condition which is not of interest. In the remaining regions, for any $\alpha$ and $\beta$ the  values of the optimisation variables that maximise the throughput can be found resorting to Lemma~\ref{lemma:dropping}. In particular:
\begin{itemize}
\item for $\mathbf x \in \mathcal G_2$: $\alpha \in [0,1]$ implies $x_1=1$ and $y_1=\alpha-1$, while $\beta \in [1,2]$ implies $x_2=\beta$ and $y_2=0$. By the last condition we can write $f(\mathbf x) = \coeffC (\alpha-1)$ so that the optimum lies in the $(\alpha,\beta$) pair that satisfies the constraint on $\dlrate$ with minimum $\alpha$. The solution follows as $\beta=1$, $\alpha= (\dlrate-\coeffB)/\coeffA$ with a  corresponding throughput $\tp=\dlrate-(\coeffC/\coeffA) (\dlrate-\tpSA)$.
\item by a symmetrical reasoning, for $\mathbf x \in \mathcal G_3$ the optimal solution is given by $\alpha=1$ and $\beta= (\dlrate-\coeffA)/\coeffB$, with a throughput $\tp=\dlrate-(\coeffD/\coeffB) (\dlrate-\tpSA)$ achieved for $x_1=1$, $y_1=0$, $x_2=1$ and $y_2=(\dlrate-\tpSA)/\coeffB$.
\item for $\mathbf x \in \mathcal G_4$: by Lemma~\ref{lemma:dropping}, $x_1=1$, $x_2=1$, so that $f(\mathbf x)=\coeffC(\alpha-1)+\coeffD(\beta-1)$. Recalling that $\beta=(\dlrate- \coeffA\alpha)/\coeffB$, we can then write $f(\mathbf x)=\alpha(\coeffC- \coeffA \coeffD/\coeffB) - \coeffC - \coeffD - \dlrate \coeffD/\coeffB$, which represents a straight line with positive slope and minimum in the left extremal point of the $\alpha$ domain. Imposing $\beta \in [1,2]$, the support of interest follows as: $\max \{ 1, (\dlrate-2 \coeffB)/\coeffA\} \leq \alpha \leq \min \{ 2, (\dlrate-\coeffB)/\coeffA\}$. Two cases have then to be distinguished. When $(\dlrate-2\coeffB)/\coeffA < 1$, $\alpha=1$ and the problem collapses to the solution found for region $\mathcal G_3$. Conversely, when $\dlrate \geq \coeffA+2\coeffB$, the optimum is achieved for $\alpha=(\dlrate-2\coeffB)/\coeffA$ and $\beta=2$, for a throughput $\tp=\dlrate-(\coeffC/\coeffA)(\dlrate-\tpSA-\coeffB)-\coeffD$ with $x_1=1$, $y_1=(\dlrate-\tpSA-\coeffB)/\coeffA$, $x_2=1$, $y_2=1$.
\end{itemize}
Comparing the  throughput of the different configurations and taking advantage of the inequality $\coeffC/\coeffA > \coeffD/\coeffB$ it is immediate to verify that the optimal solution is to pick $\mathbf x \in \mathcal G_3$ for $\dlrate\in [ \tpSA, \coeffA+2\coeffB )$ and $\mathbf x \in \mathcal G_4$ for $\dlrate\in [ \coeffA+2\coeffB, 2\tpSA ]$, stating the result of the proposition.  \hfill $\blacksquare$

\linespread{1}
\bibliographystyle{IEEEtran}
\bibliography{IEEEabrv,aloha}

\end{document}